%% file: QCD-10-002_temp.tex
\pdfoutput=1
\documentclass[11pt,twoside,a4paper,cmspaper,final,collab]{cms-tdr}
\def\svnVersion{13:17418M}\def\cmsCernNo{2010-031}\def\cmsCernDate{\today}\def\cmsMessage{Submitted to Journal of High Energy Physics}

\begin{document}\cmsNoteHeader{QCD-10-002}
%
%
%

%
%
\hyphenation{env-iron-men-tal}
\hyphenation{had-ron-i-za-tion}
\hyphenation{cal-or-i-me-ter}
\hyphenation{de-vices}
%
\RCS$Revision: 17418 $
\RCS$HeadURL: svn+ssh://alverson@svn.cern.ch/reps/tdr2/papers/QCD-10-002/trunk/QCD-10-002.tex $
\RCS$Id: QCD-10-002.tex 17418 2010-09-21 13:05:13Z alverson $
\input{ptdr-definitions}
\cmsNoteHeader{QCD-10-002} 
\title{Observation of Long-Range, Near-Side Angular Correlations in Proton-Proton Collisions at the LHC}

\address[MIT]{Massachusetts Institute of Technology}
\address[CERN]{CERN}
\author[MIT]{Wei Li, Viral Modi, Christof Roland, Gunther Roland, George Stephans, Cyrus Vafadari, Dragos Velicanu, Edward Wenger, Andre Yoon}\author[CERN]{Gabor Veres}

\date{\today}

\abstract{

Results on two-particle angular correlations for charged particles emitted in proton-
proton collisions at center-of-mass energies of 0.9, 2.36, and 7 TeV are presented, using data collected
with the CMS detector over a broad range of pseudorapidity ($\eta$) and azimuthal angle ($\phi$).
Short-range correlations in $\Delta \eta$, which are studied in minimum bias events, are characterized
using a simple ``independent cluster'' parametrization in order to quantify their strength
(cluster size) and their extent in $\eta$ (cluster decay width).
Long-range azimuthal correlations are studied differentially
as a function of charged particle multiplicity and particle transverse
momentum using a 980~nb$^{-1}$ data set at 7~TeV. In high multiplicity events, a
pronounced structure emerges in the two-dimensional correlation
function for particle pairs  with  intermediate $p_T$ of \mbox{1--3~\GeVc},
$2.0<|\Delta\eta|<4.8$ and  $\Delta\phi\approx $0.
This is the first observation of such a long-range, near-side feature
in two-particle correlation functions in $pp$ or $p\bar{p}$ collisions.
}

\hypersetup{%
pdfauthor={CMS Collaboration},%
pdftitle={Observation of Long-Range Near-Side Angular Correlations in  Proton-Proton Collisions at the LHC}, 
pdfsubject={CMS},%
pdfkeywords={CMS, physics}}

\maketitle 

\section{Introduction}

This paper presents measurements of two-particle angular correlations of charged particles
emitted in proton-proton ($pp$) collisions at center of mass energies ($\sqrt{s}$)
of 0.9, 2.36, and 7~TeV.  This first study of short- and long-range correlations in $pp$ collisions
at the LHC high energy frontier provides important information for characterizing Quantum Chromodynamics (QCD)
in this energy regime, especially the mechanism of hadronization and
possible collective effects due to the high particle densities reached in these collisions.
Multiparticle correlations in high energy collisions have been measured previously
for a broad range of collision energies and colliding systems with the goal of
understanding the underlying mechanism of particle
production
~\cite{Berger:1974vn,Morel:1974ae,Eggert:1974ek,Foa:1975eu,Ansorge:1988fg,Alver:2007wy,Alver:2008gk}.

Two related studies of angular correlations have been performed using
two-dimensional $\Delta\eta$-$\Delta\phi$ correlation functions. Here $\Delta\eta$ is the
difference in pseudorapidity $\eta$ (=$-\ln(\tan(\theta/2))$, where
$\theta$ is the polar angle relative to the beam axis) between the two particles and $\Delta\phi$ is the
difference in their azimuthal angle $\phi$ (in radians). In a first analysis, $pp$ data collected with a
minimum bias trigger at
0.9, 2.36, and 7~TeV were used to study short-range correlations ($|\Delta\eta|$ less than $\approx $2).
In a second study, the long-range structure ($2.0 < |\Delta\eta|  < 4.8$) of two-particle correlation
functions was examined as a function of charged particle multiplicity and particle transverse momentum for a large
data set at 7~TeV.

In short-range  correlations in minimum bias events,
a peak with a typical width
of about one unit in $\Delta\eta$ is observed.
A useful
way to quantify this effect is to assume
that the initial interactions emit so-called ``independent clusters'',
which subsequently decay
isotropically in their own rest frame into the observed hadrons
~\cite{Berger:1974vn,Morel:1974ae,Eggert:1974ek,Ansorge:1988fg,Alver:2007wy,Alver:2008gk}.
This simple
independent cluster model (ICM)
parametrization of the observed correlation function allows a quantitative
comparison of data and models for different collision energies and collision
systems. The observed correlation strength and extent in
relative pseudorapidity between the particles are parametrized by a
Gaussian distribution. The fitted parameters in this ansatz are the cluster multiplicity
or "size" (the average number of particles into which a cluster decays)
and the decay "width" (the spread of the daughter particles in
pseudorapidity). This ansatz is only a phenomenological parametrization
which provides no insight as to the nature of the assumed clusters nor to
the mechanisms by which clusters are formed. Relating these results to
the underlying QCD dynamics requires further modeling.

To investigate long-range azimuthal correlations ($2.0 < |\Delta\eta| < 4.8$),
a high-statistics data set of high multiplicity $pp$ events at 7~TeV was used.
In current $pp$ Monte Carlo (MC) event generators,
the typical sources of such long-range correlations are momentum conservation and
away-side ($\Delta\phi \approx \pi$) jet correlations.  Measurements at the Relativistic Heavy Ion Collider (RHIC)
have revealed that the long-range structure of two-particle angular correlation functions
is significantly modified by the presence of the hot and dense matter formed in relativistic heavy ion
collisions~\cite{Alver:2008gk}. Several novel correlation structures
over large $\Delta\eta$ were observed in azimuthal correlations
for intermediate particle transverse momenta,
$p_T \approx 1-5$~\GeVc~\cite{Alver:2009id,Abelev:2009jv}.
Since the particle densities produced in the
highest multiplicity $pp$ collisions at LHC energies begin to approach
those in high energy collisions of relatively small nuclei such as copper~\cite{xxx:2007we},
it is natural to search for the possible emergence of new features in the
two-particle correlation function from high multiplicity $pp$
events~\cite{Cunqueiro:2008uu, Luzum:2009sb, Bautista:2009my, d'Enterria:2010hd, Prasad:2009bx, Bozek:2009dt, CasalderreySolana:2009uk,Ortona:2009yc}.
Therefore, the azimuthal ($\Delta\phi$) correlation functions from
the large data set at 7~TeV have been studied differentially
by binning the events in the observed charged particle multiplicity
and by selecting particle pairs in bins of the transverse momentum of
the particles.

The paper is organized as follows: the experimental setup, event triggering, and event selection for
both analyses are described in Section~\ref{sec:exp_evt}. Criteria used to
select tracks are listed in Section~\ref{sec:track}.
The general procedure for calculating the correlation functions and the CMS-specific
efficiency corrections are described in Sections~\ref{sec:inclusive_analysis} and
\ref{sec:corrections}, respectively. Results for the analysis of short-range
correlations in minimum bias data using the cluster parametrization are given in
Section~\ref{sec:inclusive_results}. The study of long-range correlations as
a function of event multiplicity and particle transverse momentum is
detailed in Section~\ref{sec:highmult_results}.

\section{Experimental Setup, Triggering, and Event Selections}
\label{sec:exp_evt}

This analysis used three data sets collected with
$pp$ interactions at $\sqrt{s}$ = 0.9, 2.36, and 7~TeV.
A detailed description of the CMS experiment can be found in Ref.~\cite{JINST}.
The detector subsystems used for the present analysis are the pixel and
silicon-strip tracker (SST), covering the region $|\eta|< 2.5$
and immersed in a 3.8~T axial magnetic field.
The lead tungstate crystal electromagnetic calorimeter
(ECAL), the brass/scintillator hadron calorimeter (HCAL), and
the forward calorimeter (HF, covering the
region $2.9<|\eta|<5.2$), were also used for online and offline event selections.
The detailed MC simulation of the CMS detector
response is based on \GEANTfour\ \cite{GEANT4}.

Any hit in the beam scintillator counters
(BSC, $3.23<|\eta|<4.65$) coinciding with colliding proton bunches
was used for triggering the data acquisition in the minimum bias trigger.
To preferentially select non-single-diffractive (NSD) events, a
coincidence of at least one HF calorimeter tower with
more than 3 GeV of total energy on each of the positive and negative
sides was required. Events were also required to contain
at least one reconstructed primary vertex (PV) that
fell within a 4.5~cm of the nominal collision point along the beam axis
and within a radius of 0.15~cm measured perpendicular to the beam
relative to the average vertex position, and to
contain at least three fully reconstructed tracks associated with the
primary vertex. Outside this relatively narrow vertex range, the density
of events was too small to ensure enough statistics for constructing
the random background distribution (Section~\ref{sec:inclusive_analysis})
in small bins of the longitudinal ($z$) vertex position.
Beam-halo and other beam-background events were rejected
as described in Ref.~\cite{Collaboration:2010xs}. The contamination of background events
after selections in the colliding-bunch data sample was found to be negligible
($<0.1\%$).

After all selections are applied, the total number of events used for the minimum bias analysis of
cluster properties described in Section~\ref{sec:inclusive_results}
is 168\,854 ($3.3~\mu{\rm b}^{-1}$) for 0.9~TeV,
10\,902 ($0.2~\mu{\rm b}^{-1}$) for 2.36~TeV, and 150\,086 ($3.0~\mu{\rm b}^{-1}$) for 7~TeV,
where the numbers in parentheses are the approximate integrated
luminosity for the individual data samples. The systematic uncertainties in
the results shown in Section~\ref{sec:inclusive_results} significantly exceed the statistical uncertainties for
the 150k event minimum bias data sample at 7~TeV, so no further events were included
in this analysis.

In order to investigate the properties of the high multiplicity $pp$ collisions,
a dedicated high multiplicity trigger was designed and implemented in the two
levels of the CMS trigger system.
At Level 1 (L1), the
total transverse energy summed over the entire set of CMS calorimeters
(ECAL, HCAL, and HF) was required to be greater than 60~GeV. At the
high-level trigger (HLT), online tracks built from the three layers of pixel
detectors with a track origin within a cylindrical region of 21~cm along the beam
and 0.5~cm in the transverse radius were used in an online vertexing algorithm.
The number of pixel tracks (${N}_\mathrm{trk}^\mathrm{online}$) with
$|\eta|<2$, $p_{T}>0.4\GeVc$, and a distance of closest approach of 0.12~cm or less to
the best vertex (the one associated with the highest number of tracks) was determined
for each event. Data were taken with a threshold initially set to
${N}_\mathrm{trk}^\mathrm{online}>70$. During later, higher-luminosity
running, the lower limit was raised to 85.

The total integrated
luminosity for the high multiplicity analysis was 980~nb$^{-1}$.
The total number of events in each of the bins of offline reconstructed
track multiplicity, $N_\mathrm{trk}^\mathrm{offline}$,
used in the analysis are listed in Table~\ref{tab:multbinning}.
To take advantage of the full acceptance of the CMS tracking system,
$N_\mathrm{trk}^\mathrm{offline}$ includes tracks within
$|\eta|<2.4$ (see Section~\ref{sec:track} for other offline track selection criteria).
The table also lists the average values of $N_\mathrm{trk}^\mathrm{offline}$ as well
as the average of $N_\mathrm{trk}^\mathrm{corrected}$, the event multiplicity corrected
for all detector and algorithm inefficiencies, as described in Section~\ref{sec:corrections}.

\begin{table}[ht]\renewcommand{\arraystretch}{1.2}\addtolength{\tabcolsep}{-1pt}
\centering
\caption{\label{tab:multbinning} Number of events for each multiplicity bin used in the 7 TeV analysis
with total integrated luminosity of 980~nb$^{-1}$. The multiplicity of
offline reconstructed tracks,
$N_\mathrm{trk}^\mathrm{offline}$,
was counted within the kinematic cuts of $|\eta|<2.4$ and $p_{T}>0.4\GeVc$.
The last two columns list the average values of $N_\mathrm{trk}^\mathrm{offline}$
as well as the average of $N_\mathrm{trk}^\mathrm{corrected}$, the event
multiplicity corrected for all detector and algorithm inefficiencies. \\
}
\begin{tabular}{ l | l | l | l}
\hline
Multiplicity bin ($N_\mathrm{trk}^\mathrm{offline}$) & Event Count & $\left<N_\mathrm{trk}^\mathrm{offline}\right>$ & $\left<N_\mathrm{trk}^\mathrm{corrected}\right>$\\
\hline
MinBias & 21.43M & 15.9 & 17.8\\
$N_\mathrm{trk}^\mathrm{offline}<35$ & 19.36M & 13.0 & 14.1\\
$35 \leq N_\mathrm{trk}^\mathrm{offline}<90$ & 2.02M & 45.3 & 53.1\\
$90 \leq N_\mathrm{trk}^\mathrm{offline}<110$ & 302.5k & 96.6 & 111.7\\
$N_\mathrm{trk}^\mathrm{offline} \geq 110$ & 354.0k & 117.8 & 136.1\\
\hline
\end{tabular}
\end{table}

\section{Track Selection}
\label{sec:track}

In this analysis, the so-called CMS \textit{highPurity}~\cite{Khachatryan:2010pw}
tracks were used. Additionally, a reconstructed track was considered as a
primary-track candidate if the significance of the separation along the beam axis,
$z$, between the track and the primary vertex, $d_z/\sigma(d_z)$, and the significance
of the impact parameter relative to the primary vertex transverse to the beam,
$d_{\rm xy}/\sigma(d_{\rm xy})$, were each less than 3. In order to remove tracks
with potentially poorly reconstructed momentum values, the relative uncertainty
of the momentum measurement, $\sigma(p_{T})/p_{T}$, was required to be less than 10\%.

To ensure reasonable tracking efficiency and low fake rate, only tracks within $|\eta|<2.4$
and above a minimum $p_T$ value were used.
For the inclusive analysis, the selected range was
0.1~\GeVc $< p_{T} <$ 5.0~\GeVc. The effect of the upper limit imposed on $p_{T}$
is negligible. The effects of the lower $p_{T}$ cut, as well as the effect of the $\eta$ restriction
on the determination of
cluster parameters from $\Delta\eta$ correlations, are significant and
will be discussed in more detail below. To avoid possible bias in the high multiplicity
analysis, the lower cutoff was raised to $p_{T}>0.4\GeVc$ when classifying the event
multiplicity in order to match the cut applied in the online tracking.

\section{Calculation of the Two-Particle Correlation Function}
\label{sec:inclusive_analysis}

For both minimum bias and high-multiplicity triggered collision events, the first step
in extracting the correlation function was
to divide the sample into bins in track multiplicity.
For the minimum bias sample, 10 bins were used,
each containing about the same number of events.
Following an approach similar to that in Refs.~\cite{Eggert:1974ek,Alver:2007wy},
the $p_{T}$-inclusive charged two-particle correlation as a function of
$\Delta\eta$ and $\Delta\phi$ is defined as follows:

\vspace{-0.4cm}
\begin{equation}
\label{2pcorr_incl}
R(\Delta\eta,\Delta\phi) =
\left<(\left<N\right>-1)\left(\frac{S_{N}(\Delta\eta,\Delta\phi)}
{B_{N}(\Delta\eta,\Delta\phi)}-1\right)\right>_{bins}
\end{equation}

\noindent where $S_N$ and $B_N$ are the signal and
random background distributions, defined in Eqs.~(\ref{2pcorr_signal}) and
(\ref{2pcorr_background}) respectively,
$\Delta\eta(=\eta_1-\eta_2)$ and $\Delta\phi(=\phi_1-\phi_2)$ are the
differences in pseudorapidity and azimuthal angle between the two particles,
$\left<N\right>$ is the number of tracks per event averaged over the multiplicity bin,
and the final $R(\Delta\eta,\Delta\phi)$ is found by averaging over multiplicity bins.
For simplicity in Eq.~(\ref{2pcorr_incl}) and the discussion in this section, $N$ is used to
represent the total number of offline reconstructed tracks per event.
Note that the order in which the particles are considered has no significance. The quantities $\Delta\eta$ and
$\Delta\phi$ are always taken to be positive and used to fill one quadrant of
the $\Delta\eta,\Delta\phi$ histograms with the other three quadrants filled by
reflection. Therefore, the resulting distributions are symmetric about
($\Delta\eta$,$\Delta\phi$)=(0,0) by construction.

For each multiplicity bin,
the signal distribution:

\vspace{-0.4cm}
\begin{equation}
\label{2pcorr_signal}
S_{N}(\Delta\eta,\Delta\phi)=\frac{1}{N(N-1)}
\frac{d^{2}N^{\rm signal}}{d\Delta\eta d\Delta\phi}
\end{equation}

\noindent was determined by counting all particle pairs within each event,
using the weighting factor $N(N-1)$,
then averaging over all events.
This represents the charged two-particle pair density function normalized to unit integral.
The background distribution:

\vspace{-0.4cm}
\begin{equation}
\label{2pcorr_background}
B_{N}(\Delta \eta,\Delta \phi)=\frac{1}{N^{2}}
\frac{d^{2}N^{\rm mixed}}{d\Delta\eta d\Delta\phi}
\end{equation}

\noindent denotes the distribution of uncorrelated particle pairs representing a product of two single-particle
distributions, also normalized to unit integral.
This distribution was constructed by randomly selecting two different events
within the same multiplicity bin and pairing every particle from one
event with every particle in the other (in this case, the normalization factor $1/N^{2}$ corresponds
to $1/N_1N_2$ event-by-event).
The pairs of events used to compute the background were also required to be within the same
0.5~cm wide bin in the vertex location along the beam.

As indicated in Eq.~(\ref{2pcorr_incl}),
the ratio of $S_{N}(\Delta\eta,\Delta\phi)$
to $B_{N}(\Delta \eta,\Delta \phi)$ was first calculated in each
multiplicity bin. Dividing the background in this way
corrects for detector effects such as tracking inefficiencies, non-uniform acceptance, etc.
The ratio of signal to background was then weighted
by the track multiplicity factor, $\left<N\right>-1$ (where $\left<N\right>$ is the average multiplicity in each bin),
and averaged over all the multiplicity bins to arrive
at the final two-particle correlation function $R(\Delta\eta,\Delta\phi)$.

\section{Corrections for Tracking and Event Selection Inefficiencies}
\label{sec:corrections}

\subsection{Correction for Tracking Inefficiency}

Studies with simulated events showed that the combined geometrical acceptance
and reconstruction efficiency for the global track reconstruction exceeds 50\%
around $p_{T}$ $\approx$ 0.1~\GeVc over the full CMS tracker acceptance ($|\eta|<2.4$) for charged hadrons.
The efficiency is greater than
90\% in the $|\eta|<1$ region for $p_{T} >$ 0.6~\GeVc. Detailed studies of tracking
efficiencies using MC-based and data-based methods can be found in \cite{TRK-10-002}. The tracking efficiency
correction factor, $\epsilon^{\mathrm{trk}}$, was determined by taking the ratio of the number
of reconstructed tracks ($N^{\rm trk}$) to that of generator level primary charged particles ($N^{\rm gen}$) in
the simulated MC events as a function of $p_{T}$, $\eta$, $z_{\rm vtx}$:

\vspace{-0.4cm}
\begin{equation}
\label{trackingefficiency}
\epsilon^{\mathrm{trk}}(\eta,p_{T},z_{\rm vtx}) =
\frac{N^{\rm trk}(\eta,p_{T},z_{\rm vtx})}{N^{\rm gen}(\eta,p_{T},z_{\rm vtx})}.
\end{equation}

In constructing the signal and background distributions,
this correction was applied as an inverse weight, $1/\epsilon^{\rm trk}(\eta,p_{T},z_{\rm vtx})$,
to each particle.
After this correction, the two-particle correlation function found using reconstructed
tracks from simulated events matched that obtained
at the generator level to within 1.4\%.

Using simulations, the tracking efficiency was found to have little or no dependence on
multiplicity within the range studied in the present work. The fake rate did increase
slightly with multiplicity but remained at the 1-2\% level. Therefore, the corrections
applied for tracking efficiency and fake rate were independent of event multiplicity.

\subsection{Event Selection Correction for Minimum Bias Data}
\label{sec:inclusive_evtcorr}

For the minimum bias data, inefficiencies in triggering and vertex reconstruction of low
multiplicity events resulted in multiplicity distributions of reconstructed tracks which
were biased toward higher average values. The correction factor for this effect,
$\epsilon^{evt}$, was determined by taking the ratio of two generator-level MC multiplicity
distributions, one with offline event selection applied ($N^{\rm evtSel}_{\rm gen}$)
and one for all NSD MC events ($N^{\rm NSD}_{\rm gen}$).
In Eq.~(\ref{eventefficiency}), $N_{\rm trk}^{\rm true}$ represents
the true number of particles in the event. The NSD event selection efficiency:

\vspace{-0.4cm}
\begin{equation}
\label{eventefficiency}
\epsilon^{evt}(N_{\rm trk}^{\rm true}) =
\frac{N^{\rm evtSel}_{\rm gen}(N_{\rm trk}^{\rm true})}{N^{\rm NSD}_{\rm gen}(N_{\rm trk}^{\rm true})}
\end{equation}

is about 50\% at $N_{\rm trk}^{\rm true}$=6 and reaches 100\%
around $N_{\rm trk}^{\rm true}$=15.
When calculating the correlation function, each event was weighted
by the inverse of the event selection efficiency evaluated at
$N_\mathrm{trk}^\mathrm{corrected}$ which is the number of particles corrected for acceptance and
tracking efficiency as described above,
$1/\epsilon^{evt}(N_\mathrm{trk}^\mathrm{corrected})$.

\subsection{Event Selection Correction for High Multiplicity Data}
\label{sec:highmult_evtcorr}

\begin{figure}[ht!]
\centerline{
  \mbox{\includegraphics[width=0.85\linewidth]{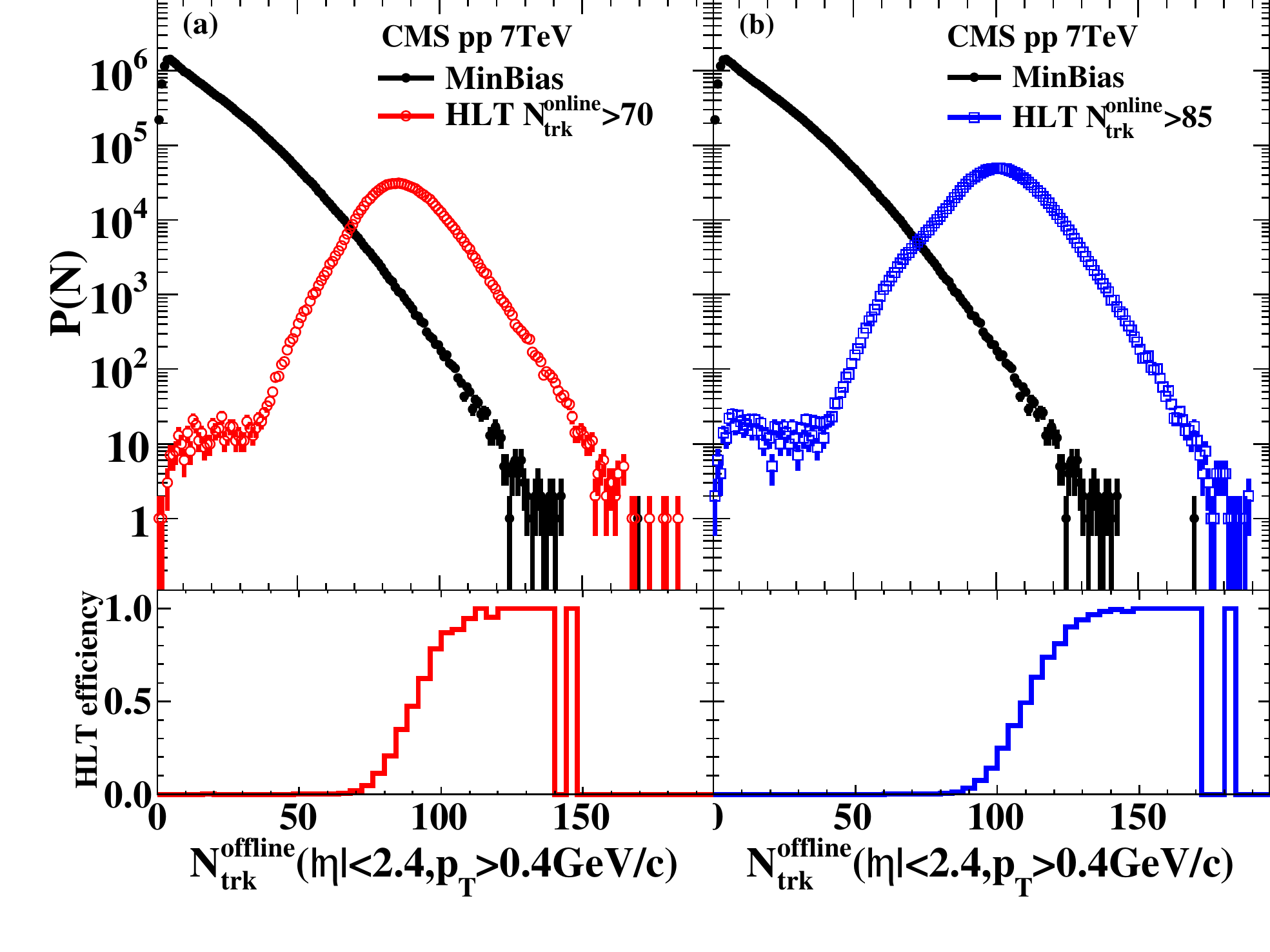}}}
\vspace{-0.03cm}
  \caption{ \label{fig:Multiplicity_ALL} Distributions of the number of tracks reconstructed
  in offline analysis, $N_\mathrm{trk}^\mathrm{offline}$,
  for minimum
  bias events, as well as high-multiplicity triggered events, both at 7~TeV,  with online multiplicity
  $N_\mathrm{trk}^\mathrm{online}$ greater than
  (a) 70 and (b) 85. The total integrated luminosity of the data set is
  980~nb$^{-1}$. The minimum bias trigger was heavily prescaled during higher
  luminosity LHC running. The HLT efficiency turn-on curves
  for the two high multiplicity triggers are shown in the two panels at the bottom.
   }
\end{figure}

The two high-multiplicity trigger thresholds used in the HLT (see Section~\ref{sec:exp_evt})
give different trigger efficiencies.
Distributions for offline reconstructed track multiplicity,
$N_\mathrm{trk}^\mathrm{offline}$, in minimum bias and high
multiplicity triggered events at 7~TeV are shown in the top panels of Fig.~\ref{fig:Multiplicity_ALL}.
Kinematic cuts of
$|\eta|<2.4$ and $p_{T}>0.4\GeVc$ were used in defining $N_\mathrm{trk}^\mathrm{offline}$
(see Section~\ref{sec:track} for other offline track selections).
The statistics of events with $N_\mathrm{trk}^\mathrm{offline} \geq 110$ were enhanced by a factor of about 1000
with the high multiplicity trigger relative to the minimum bias trigger due the large prescale
factor applied to the latter sample.
The lower panels in Fig.~\ref{fig:Multiplicity_ALL} show the HLT efficiency, obtained from data, of the two
high multiplicity triggers relative to the minimum bias trigger.
The L1 triggering efficiency (not shown in Fig.~\ref{fig:Multiplicity_ALL}) is not
a concern since it reaches 100\% efficiency for  events with
$N_\mathrm{trk}^\mathrm{offline}\geq 90$.
A weight given by the inverse of the HLT efficiency,
$\varepsilon^{HLT}_{evt}(N_\mathrm{trk}^\mathrm{offline})$, was applied to all pairs from
a given event.

\section{Short-Range Correlations in 0.9, 2.36, and 7~TeV Data}
\label{sec:inclusive_results}

\begin{figure}[thb!]
\centerline{
  \mbox{\includegraphics[width=\linewidth]{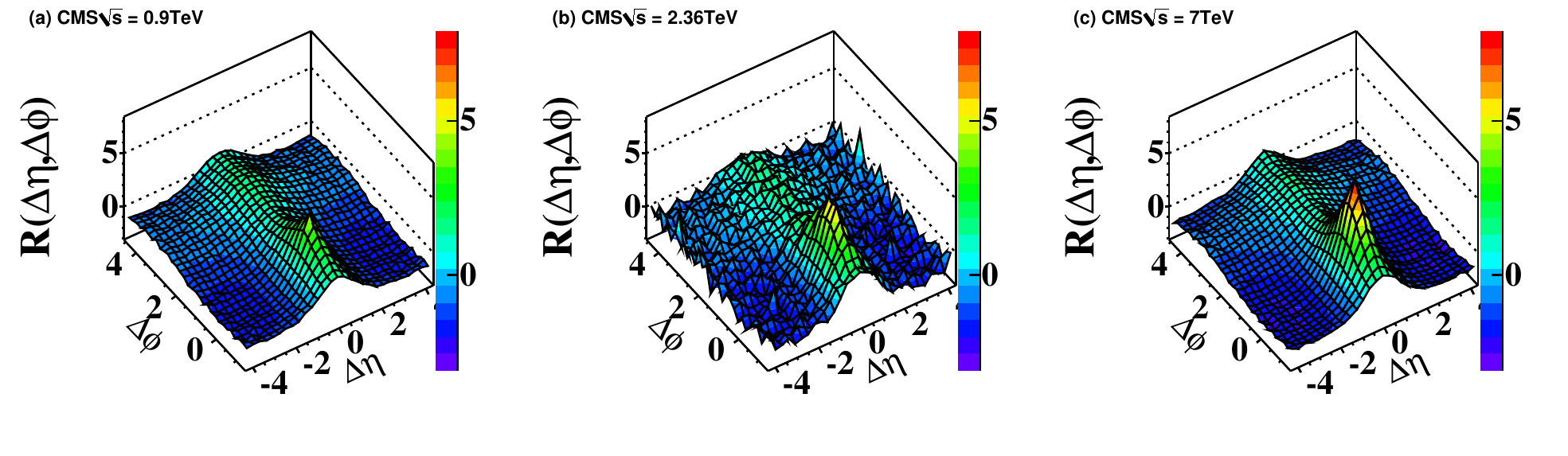}}}
\vspace{-0.3cm}
  \caption{ \label{corr_2D_finaldata_bothenergy}
  Two-particle correlation functions versus $\Delta \eta$ and $\Delta \phi$
  in $pp$ collisions at $\sqrt{s} =$ (a) 0.9, (b) 2.36, and (c) 7~TeV. }
\end{figure}

\begin{figure}[thb!]
\centerline{
  \mbox{\includegraphics[width=\linewidth]{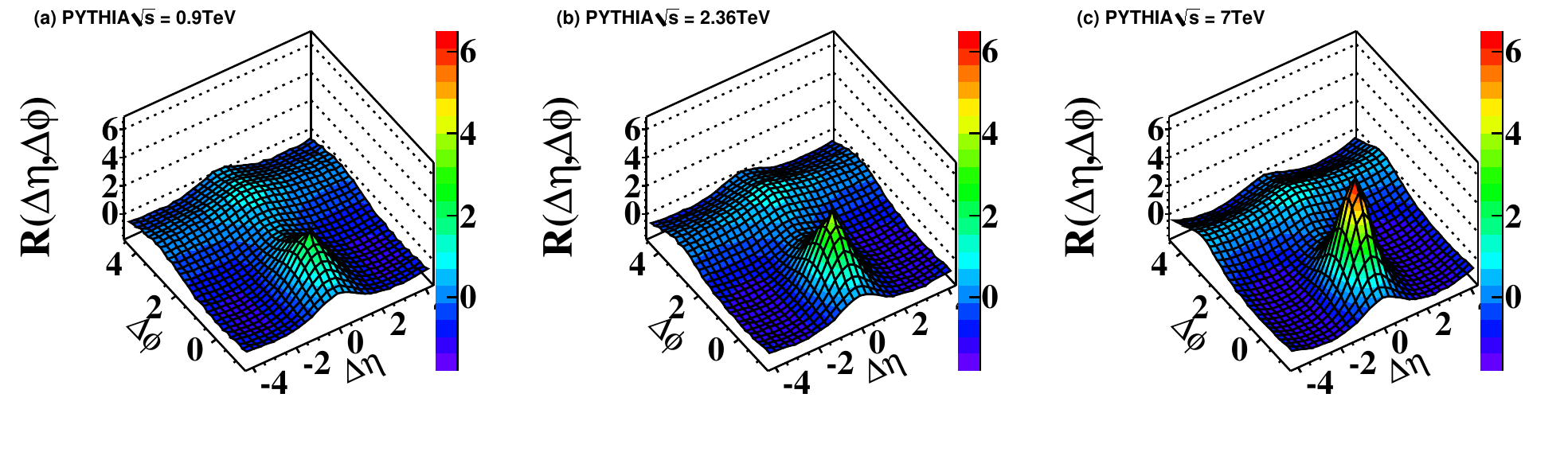}}}
\vspace{-0.3cm}
  \caption{ \label{corr_2D_mc_bothenergy}
  Two-particle correlation functions versus $\Delta \eta$ and $\Delta \phi$
  in PYTHIA D6T tune at $\sqrt{s} =$ (a) 0.9, (b) 2.36, and (c) 7~TeV. }
\end{figure}

The final two-particle inclusive correlation functions are shown in
Fig.~\ref{corr_2D_finaldata_bothenergy} as a function
of $\Delta \eta$ and $\Delta \phi$ at $\sqrt{s}$ = 0.9, 2.36, and 7~TeV.
A small region with $|\Delta\eta|<0.06$ and $|\Delta\phi|<0.06$ was excluded
in both signal and background distributions in order to reject residual secondary effects
(i.e., any tracks from photon conversions, weak decays, or
$\delta$-electrons which were not rejected by the cut on the projected distance of the track
from the vertex).

The complex two-dimensional (2-D) correlation structure shown in Fig.~\ref{corr_2D_finaldata_bothenergy}
is dominated by three prominent components: a narrow peak at
($\Delta\eta$,$\Delta\phi$)$ \approx $(0,0)
which can be understood as the contribution from higher $p_{T}$
clusters (e.g., hard processes like jets); a ridge at $\Delta\phi \approx \pi$ spread over a
broad range in $\Delta\eta$, interpreted as due to away-side jets or more generally
momentum conservation; and an approximately Gaussian ridge at $\Delta\eta \approx $0
extending over the whole range of $\Delta\phi$,
becoming broader toward larger $\Delta \phi$ values,
which arises
from the decay of clusters with lower $p_{T}$ (e.g., soft QCD string fragmentation).
This broadening will be discussed in quantitative
detail later in this section. The PHOBOS experiment at RHIC observed similar correlation structures
in $pp$ collisions at $\sqrt{s}$ = 200~GeV and 410~GeV~\cite{Alver:2007wy}.
Qualitatively similar structures also exist in PYTHIA
(Fig.~\ref{corr_2D_mc_bothenergy} for D6T tune~\cite{Bartalini:2009xx})
although they do not reproduce the strength of the correlations seen in the data.
The qualitative features of the observed correlations in the data are also consistent with
an independent cluster approach according to a simulation study from the ISR experiment
using a low-mass resonance ($\rho$, $\omega$, $\eta$) gas model~\cite{Eggert:1974ek}
and a MC model of isotropic cluster decays from the PHOBOS experiment~\cite{Alver:2008gk}.
Bose--Einstein Correlations (BEC, also known as the Hanbury-Brown and Twiss effect~\cite{Lisa:2008gf})
have been measured in $pp$ collisions~\cite{Albajar:1989sj, Khachatryan:2010un, Aamodt:2010jj} but
their influence on the extracted cluster parameters has been found to be
negligible \cite{Alver:2007wy}.

To quantify one aspect of the correlation structure, the 2-D correlation functions were reduced to
one-dimensional (1-D) functions of $\Delta \eta$ by integrating
$S_{N}(\Delta \eta,\Delta \phi)$ and $B_{N}(\Delta \eta,\Delta \phi)$ over $\Delta \phi$:

\vspace{-0.4cm}
\begin{equation}
\label{2pcorr_1Dprojection}
R(\Delta \eta)=\left<(\left<N\right>-1)\left(\frac{\int S_{N}
(\Delta \eta, \Delta \phi) d\Delta \phi}{\int B_{N}
(\Delta \eta, \Delta \phi) d\Delta \phi}-1\right)\right>_{bins}.
\end{equation}

\begin{figure}[bht!]
\centerline{
  \mbox{\includegraphics[width=\linewidth]{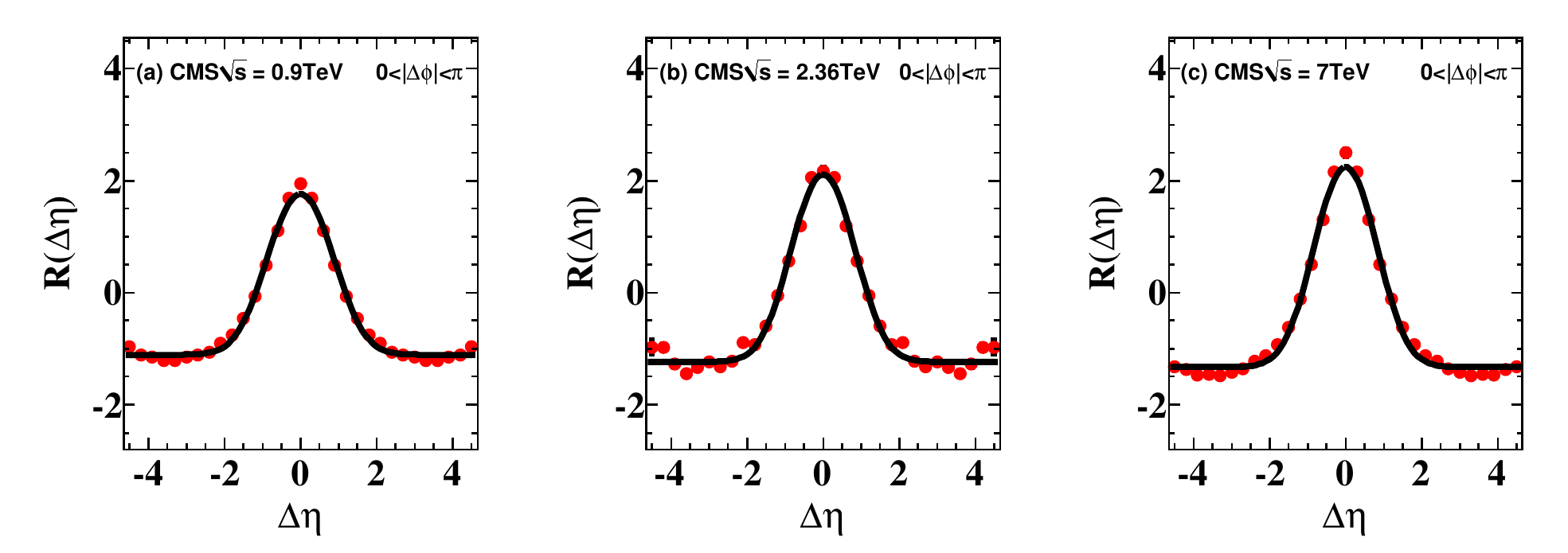}}}
\vspace{-0.3cm}
  \caption{ \label{2pcorr_eta}
  Two-particle pseudorapidity correlation function, obtained by averaging over the
  entire $\Delta \phi$ range from $0$ to $\pi$, in $pp$ collisions
  at $\sqrt{s} =$ (a) 0.9, (b) 2.36, and (c) 7~TeV.
  The solid curves correspond to the fits by the cluster model
  using Eq.~(\ref{2pcorr_clusterfitting_incl}).
  Error bars are smaller than the symbols.
}
\end{figure}

\begin{figure}[ht]
\centerline{
  \mbox{\includegraphics[width=0.5\linewidth]{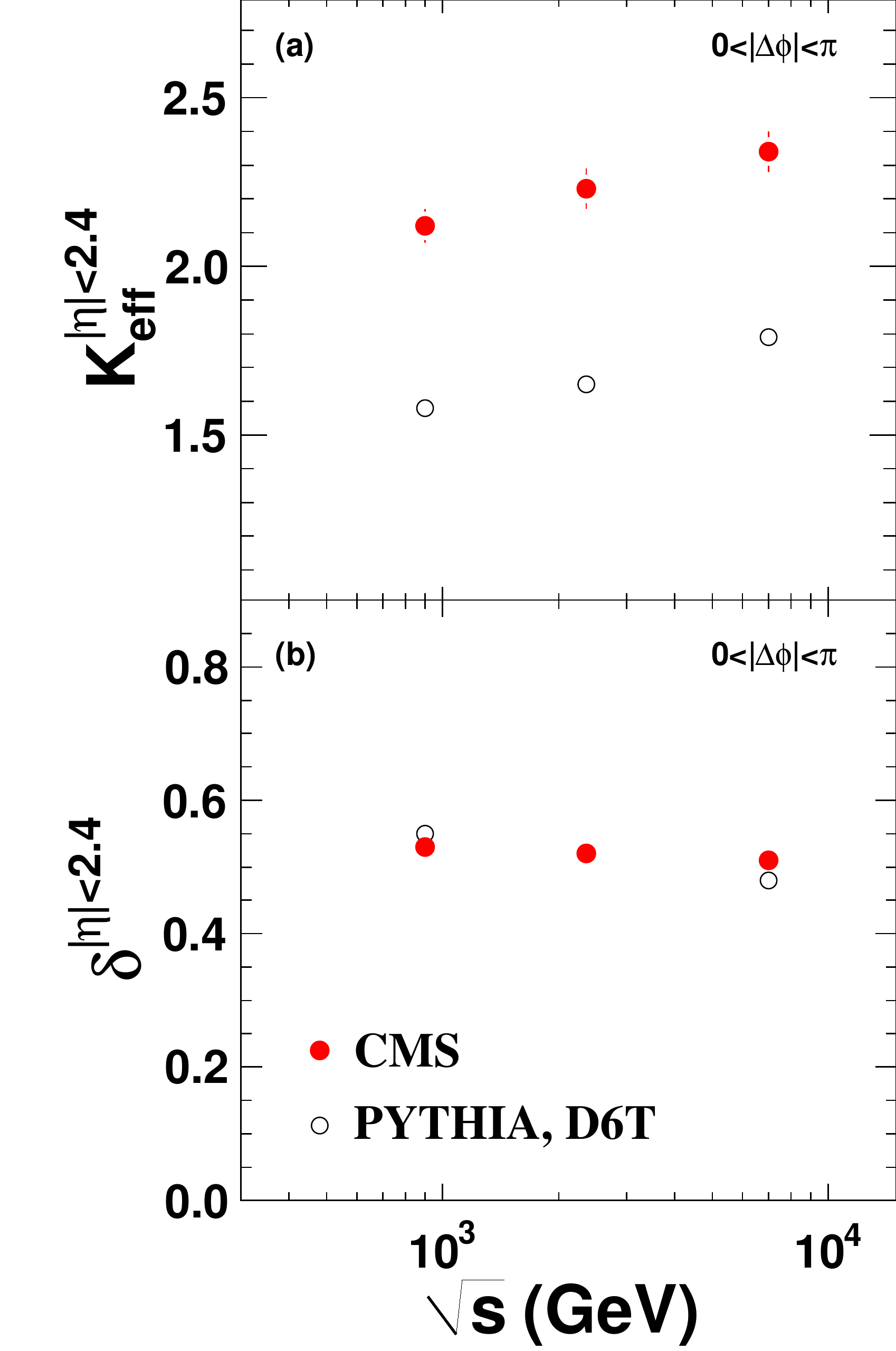}}}
\vspace{-0.3cm}
  \caption{ (a) $K_{\rm eff}$ and (b) $\delta$ as a function of $\sqrt{s}$,
            measured for $p_{T}>0.1\GeVc$ and $|\eta|<2.4$ by CMS in solid circles. Open circles
            show the PYTHIA results with the D6T tune.}
  \label{cluster_sqrts_cms}
\end{figure}

\noindent The 1-D two-particle pseudorapidity correlation functions, $R(\Delta \eta)$,
where $\Delta \phi$ was averaged over the entire range from 0 to $\pi$,
are shown for all three energies in Fig.~\ref{2pcorr_eta}.

In the context of an ICM description, $R(\Delta \eta)$
can be parametrized using the functional form~\cite{Morel:1974ae}:

\vspace{-0.4cm}
\begin{equation}
\label{2pcorr_clusterfitting_incl}
R(\Delta \eta)=\alpha\left[\frac{\Gamma(\Delta \eta)}{B(\Delta \eta)}-1\right]
\end{equation}

\noindent where the correlation strength $\alpha=\frac{ \langle K(K-1)
\rangle }{ \langle K \rangle }$ depends on the average numbers of particles into which
a cluster decays, the cluster size $K$. The function $\Gamma(\Delta \eta)$ is
a Gaussian function proportional to \[ \exp{[- (\Delta \eta)^{2}/(4\delta^{2})]} \]
where $\delta$ quantifies the average spread of particles originating from a single
cluster, i.e.\ the decay width.
The background distribution, $B(\Delta \eta)$, in Eq.~(\ref{2pcorr_clusterfitting_incl}) is the
same event-mixed distribution defined in Eq.~(\ref{2pcorr_background}) but averaged over all the multiplicity
bins with all corrections applied, and integrated over $\Delta\phi$.

Without knowing $\sigma_K$, the width of the distribution of $K$, it is impossible to
calculate the average cluster size $\langle K \rangle$ directly from the measured
value of $\alpha$.
However, an effective cluster size can be defined using the extracted correlation strength via the relation:

\vspace{-0.4cm}
\begin{equation}
\label{Keff}
K_{\rm eff}=\alpha+1=\frac{\left<K(K-1)\right>}{\left<K\right>}+1=\left<K\right>+\frac{\sigma_{K}^{2}}{\left<K\right>}.
\end{equation}

\noindent
The effective cluster size $K_{\rm eff}$
and decay width $\delta$ can be estimated by means of a least $\chi^{2}$ fit
of Eq.~(\ref{2pcorr_clusterfitting_incl}) to the measured two-particle pseudorapidity
correlation function. The ICM provides a
good fit to the data over a large range in $\Delta \eta$, as shown in Fig.~\ref{2pcorr_eta}.

The statistical uncertainties of the fit parameters are much smaller than the systematic ones.
The correction for event selection efficiency (see Section~\ref{sec:inclusive_evtcorr})
has an overall systematic uncertainty of less
than 2.8\% found by comparing the result at the generator level to that from the reconstructed
tracks after corrections. The model dependence of this procedure (i.e.\ the selection efficiency
for NSD events) was investigated by using correction factors derived
from different MC generators such as PYTHIA, PHOJET~\cite{Bopp:1998rc},
and HERWIG++~\cite{Bahr:2008pv}. The biggest discrepancy in the final results
was about 2.6\%.

Systematic uncertainties due to track quality cuts were examined by loosening the
cuts on the significance of both the transverse track impact parameter,
$d_{xy}/\sigma(d_{xy})$,
and the distance along the beam to the primary vertex, $d_z/\sigma(d_z)$ from 3 to 5. The final
results were found to be insensitive to these track selections to within 1.2\%.

A summary of systematic uncertainties for the inclusive analysis is given in
Table~\ref{tab:syst-table}. The uncertainties are presented for the cluster model fit
parameters listed in Table~\ref{tab:syst-table}, namely the correlation strength
($\alpha=K_{\rm eff}-1$) and the width in pseudorapidity ($\delta$).

\begin{table}[ht]
\caption{\label{tab:syst-table} Summary of systematic uncertainties in the inclusive analysis.}

\begin{center}
\begin{tabular}{lcc}
\hline
\hline
&\multicolumn{2}{c}{Systematic uncertainties [\%]} \\
\cline{1-3}
 Source                                          & $K_{\rm eff}-1$ ($\alpha$)& $\delta$  \\
\hline
 Track quality cuts                                                   & 1.2 & 1.0 \\
 Correction for tracking/acceptance efficiency and fake rate           & 1.3 & 1.4 \\
 Correction for event selection efficiency                             & 2.6 & 2.8 \\
 Model dependence of the corrections                                  & 2.6 & 1.3 \\
\hline
 Total systematic uncertainties                                       & 4.1 & 3.5\\
\hline
\hline
\end{tabular}
\end{center}
\end{table}

Values of effective cluster sizes and widths observed within the kinematic cuts on
$p_{T}$ and $|\eta|$ are summarized in Table~\ref{table:results_cms}.

\begin{table}[ht]
\caption{\label{table:results_cms} Final results on $K_{\rm eff}$ and $\delta$ measured
within the kinematic cuts of $p_{T}>0.1\GeVc$ and $|\eta|<2.4$ at CMS.}

\begin{center}
\begin{tabular}{lcc}
\hline
\hline
 $\sqrt{s}$                       & $K_{\rm eff}^{|\eta|<2.4}$ & $\delta^{|\eta|<2.4}$ \\
\hline
 0.9~TeV                          & $2.12 \pm <0.01 \stat \pm 0.05 \syst$ & $0.53 \pm <0.01 \stat \pm 0.02 \syst$ \\
 2.36~TeV                         & $2.23 \pm 0.02 \stat \pm 0.05 \syst$ & $0.52 \pm <0.01 \stat \pm 0.02 \syst$ \\
 7~TeV                            & $2.34 \pm <0.01 \stat \pm 0.06 \syst$ & $0.51 \pm <0.01 \stat \pm 0.02 \syst$ \\
\hline
\hline
\end{tabular}
\end{center}
\end{table}

As can be seen in Fig.~\ref{2pcorr_eta}, the most central point of the 1-D pseudorapidity correlation function always lies slightly
above the fits. This could be due to the residual effects
of secondary processes that were not fully removed by the track selection,
as well as BEC
or other physics processes at this small scale in $\Delta \eta$. All fits exclude this
central point, but including it in the fit affects the values of $K_{\rm eff}$ and $\delta$ by no more than 0.5\%.

In Fig.~\ref{cluster_sqrts_cms}, CMS measurements of $K_{\rm eff}$ and $\delta$
for $p_{T}>0.1\GeVc$ and $|\eta|<2.4$ are shown as functions of $\sqrt{s}$, and
compared with the PYTHIA D6T tune.
An energy dependence of $K_{\rm eff}$ is observed,
while $\delta$ remains roughly constant over the three energies.
PYTHIA shows energy dependencies of $K_{\rm eff}$ and
$\delta$ similar to those seen in the data, but the magnitude of $K_{\rm eff}$
is  significantly smaller in PYTHIA.
The effect of tensor mesons in PYTHIA was investigated, but
even using an unrealistically large probability
of 50\% for the angular momentum L=1 meson states accounted for only about one
third of the difference.
Results from the HERWIG++ model were also studied and found to have correlation function shapes
dramatically different from the data, in agreement with previous results~\cite{Abdallah:2002mk}
showing that HERWIG++ is insufficiently tuned to reproduce soft QCD processes.

\begin{table}[th]
\caption{\label{table:results_extra} Final results on $K_{\rm eff}$ and $\delta$ measured by CMS
after extrapolation to $p_{T} > 0$ and $|\eta|<3$. The third quoted uncertainty is due to the extrapolation procedure.}

\begin{center}
\begin{tabular}{lcc}
\hline
\hline
 $\sqrt{s}$                       & $K_{\rm eff}^{|\eta|<3.0}$ & $\delta^{|\eta|<3.0}$ \\
\hline
 0.9~TeV                          & $2.50 \pm <0.01 \stat \pm 0.06 \syst \pm 0.07$ & $0.64 \pm <0.01 \stat \pm 0.02 \syst \pm 0.03$ \\
 2.36~TeV                         & $2.65 \pm 0.03 \stat \pm 0.07 \syst \pm 0.08$ & $0.60 \pm 0.01 \stat \pm 0.02 \syst \pm 0.03$ \\
 7~TeV                            & $2.75 \pm <0.01 \stat \pm 0.07 \syst \pm 0.09$ & $0.59 \pm <0.01 \stat \pm 0.02 \syst \pm 0.03$ \\
\hline
\hline
\end{tabular}
\end{center}
\end{table}

In order to compare with measurements made at lower energies, CMS results
were extrapolated to $|\eta|<3$ and the full $p_{T}$ range $p_{T} >  0$ to achieve a consistent
kinematic range. The fraction of tracks below
$p_{T} \approx 0.1 \GeVc$ was estimated by fitting the measured $p_{T}$ distributions
using the Tsallis function as was done in Ref.~\cite{Collaboration:2010xs},
which empirically describes both the low $p_{T}$ exponential and the high $p_{T}$ power-law
behaviors~\cite{Tsallis:1987eu}. The integral of the fit function for $p_{T} < 0.1 \GeVc$
amounts to about 5.5\% of the total yield, consistent with the results in Ref.~\cite{Collaboration:2010xs}.
As first quantified in Ref.~\cite{Alver:2008gk}, the loss of particles falling outside a
limited $\eta$ acceptance results in a significant reduction of both $K_{\rm eff}$ and $\delta$.
This effect was investigated using several dynamical models as well as the simple
ICM following the identical approach used in Ref.~\cite{Alver:2008gk}.
As was the case in the previous analysis, the ratios of $K$ and $\delta$ for different
$\eta$ acceptances ($|\eta|<3.0$ and $|\eta|<2.4$ in the present work) were found to scale
very closely with $\delta^{|\eta|<2.4}$, the measured cluster width using data in
the $|\eta|<2.4$ region, reinforcing the conclusion that the dependence of the
extracted cluster parameters on pseudorapidity acceptance is primarily a simple
geometric effect.

\begin{figure}[t!]
\centerline{
  \mbox{\includegraphics[width=0.5\linewidth]{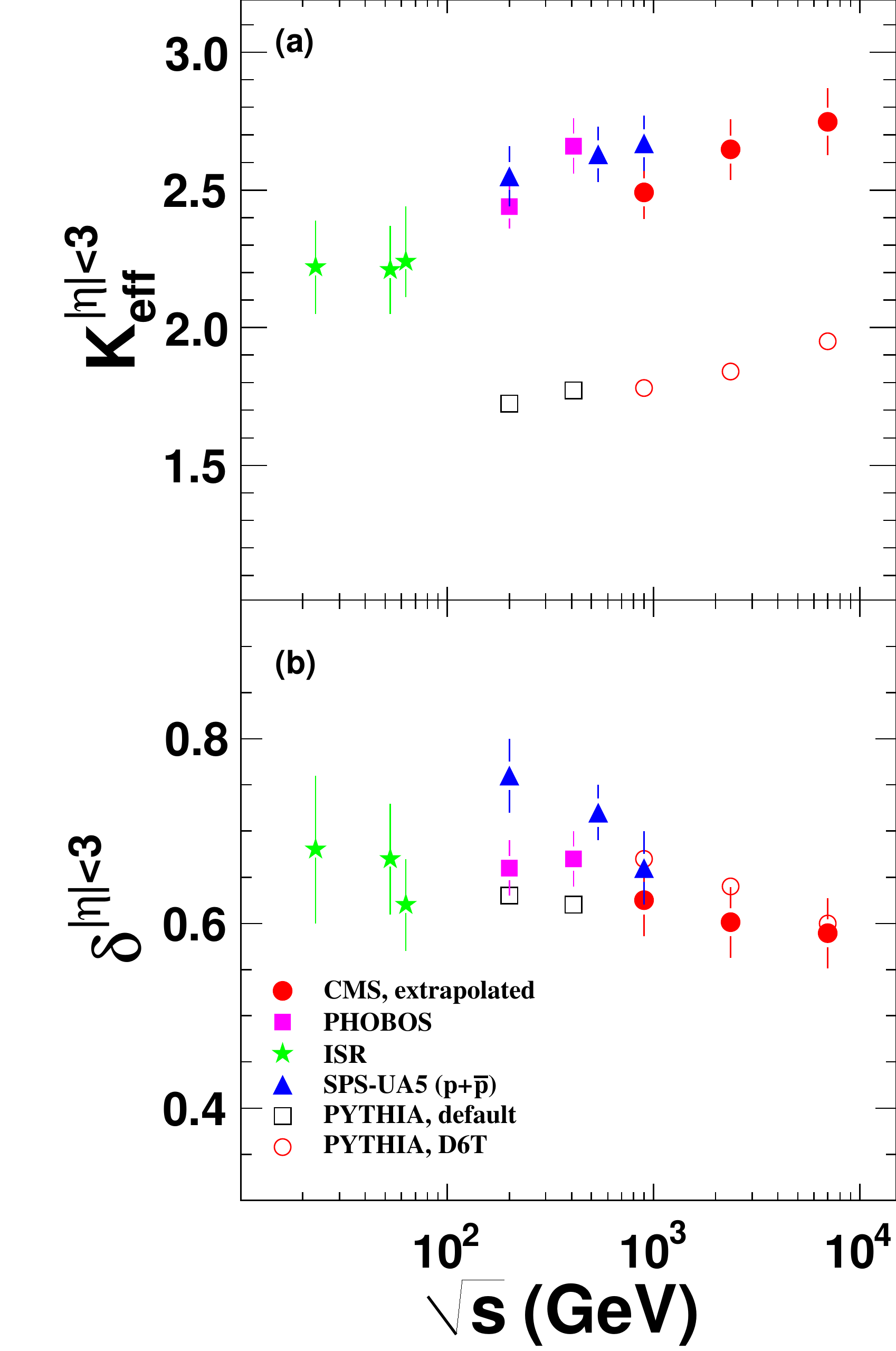}}}
\vspace{-0.3cm}
  \caption{ (a) $K_{\rm eff}$ and (b) $\delta$ as a function of $\sqrt{s}$ based on a
            model-dependent extrapolation of CMS data to $p_T\approx 0$ and $|\eta|<3$ (solid circles),
            as well as data from PHOBOS~\cite{Alver:2007wy} (solid squares), UA5~\cite{Ansorge:1988fg}
            (solid triangles) and ISR \cite{Eggert:1974ek} (solid stars)
            experiments for $pp$ and $p\bar{p}$ collisions. Open circles and squares
            show the PYTHIA results for the D6T tune and default parameters, respectively. }
  \label{cluster_sqrts}
\end{figure}

Figure~\ref{cluster_sqrts} shows the results of $K_{\rm eff}$ and $\delta$
measured by the CMS experiment after the extrapolation to $|\eta|<3$ and $p_{T} \approx 0$,
as well as previous measurements at lower energies in the same pseudorapidity
range \cite{Eggert:1974ek,Ansorge:1988fg,Alver:2007wy}.
Values of the extrapolated CMS results are summarized in Table~\ref{table:results_extra}, where the
third quoted uncertainty is due to the extrapolation.
The error bars in Fig.~\ref{cluster_sqrts} include the systematic uncertainties from both
the experimental measurements and the extrapolations added in quadrature.
Events generated with PYTHIA D6T tune show a similar energy dependence of $K_{\rm eff}$ and $\delta$
as the data, but systematically underestimate the magnitude of $K_{\rm eff}$ over the full energy range.

The observed cluster size cannot be fully explained by a resonance decay
model even at very low energies, since the expectation of $\langle K \rangle$
from resonance decays is about 1.5 (extrapolating to 1.7 for $K_{\rm eff}$
depending on the assumed cluster size distribution \cite{Ansorge:1988fg}).
This is significantly lower than the observed values, but is close to what
is seen in PYTHIA. Additional sources of pseudorapidity correlations,
such as local quantum number conservation \cite{Porter:2005rc}, are
needed to describe the data. As the energy increases further (especially at the TeV scale),
the onset of jets should play a more important role in the particle production, resulting
in bigger clusters.  This effect could be the underlying cause for the observed energy
dependence of $K_{\rm eff}$.

\section{Long-Range Correlations in 7~TeV Data}
\label{sec:highmult_results}

The study of long-range azimuthal correlations involved
generating 2-D $\Delta \eta$-$\Delta \phi$ distributions in bins of event multiplicity
and particle transverse momentum. The analysis procedure was to a large extent identical with
that used for the minimum bias data described in Section~\ref{sec:inclusive_analysis}.
With the addition of $p_T$ binning, both particles in the pairs used to calculate
$R(\Delta\eta, \Delta\phi)$ were required to be within the selected $p_T$ range.
The events were divided into bins of offline track multiplicity as outlined
in Table~\ref{tab:multbinning}. In order to reach good statistics
for the highest attainable charged particle densities, only data at 7~TeV were considered.

\begin{figure}[thb]
\centerline{
  \mbox{\includegraphics[width=0.8\linewidth]{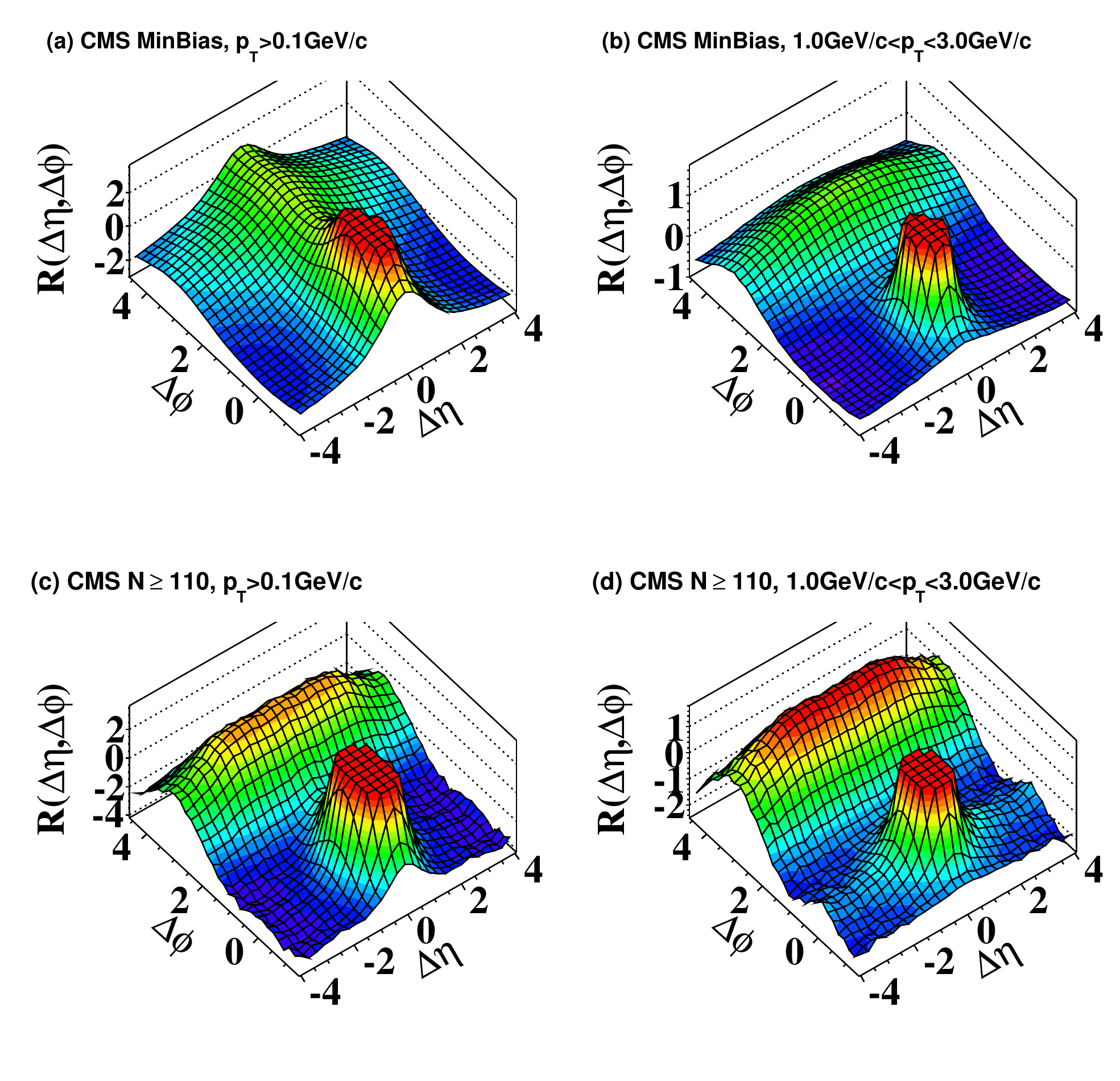}}}
\vspace{-0.3cm}
  \caption{ \label{fig:compare_corr_2D_PPData_Minbias_7TeV_all} 2-D two-particle correlation
functions for 7~TeV $pp$ (a) minimum bias events with $p_{T}>0.1\GeVc$, (b) minimum bias events with
$1 < p_{T} < 3\GeVc$, (c) high multiplicity ($N_\mathrm{trk}^\mathrm{offline} \geq 110$) events with $p_{T}>0.1\GeVc$
and (d) high multiplicity ($N_\mathrm{trk}^\mathrm{offline} \geq 110$) events with $1<p_{T}<3\GeVc$.
The sharp near-side peak from jet correlations is cut off in order to better illustrate the
structure outside that region.
   }
\end{figure}

Figure~\ref{fig:compare_corr_2D_PPData_Minbias_7TeV_all} compares 2-D two-particle correlation
functions for minimum bias events and high multiplicity events, for both inclusive particles and for
particles in an intermediate $p_{T}$ bin. The top two panels show results from minimum bias
events. The correlation function for inclusive particles with $p_{T} > 0.1$~\GeVc shows
the typical structure as described by the independent
cluster model. The region at $\Delta\eta\approx $0 and intermediate $\Delta\phi$
is dominated by particle emission from clusters with low transverse momentum, with some
contribution from jet-like particle production near $(\Delta\eta, \Delta\phi) \approx (0, 0)$
due to near-side jet fragmentation and a broad elongated ridge around $\Delta\phi
\approx  \pi$ due to fragmentation of back-to-back jets. Also visible is a shallow minimum at
$\Delta\phi \approx 0$ at large $|\Delta\eta| $ due to momentum conservation.
For the intermediate $\pt$ region of $1\GeVc<p_{T}<3\GeVc$ a more pronounced near-side
jet peak and away-side ridge are visible, due to the enhanced contribution of jet fragmentation
to particle production for increasing $p_T$.

\begin{figure}[thb]
\centerline{
\vspace{0.1cm}
  \mbox{\includegraphics[width=\linewidth]{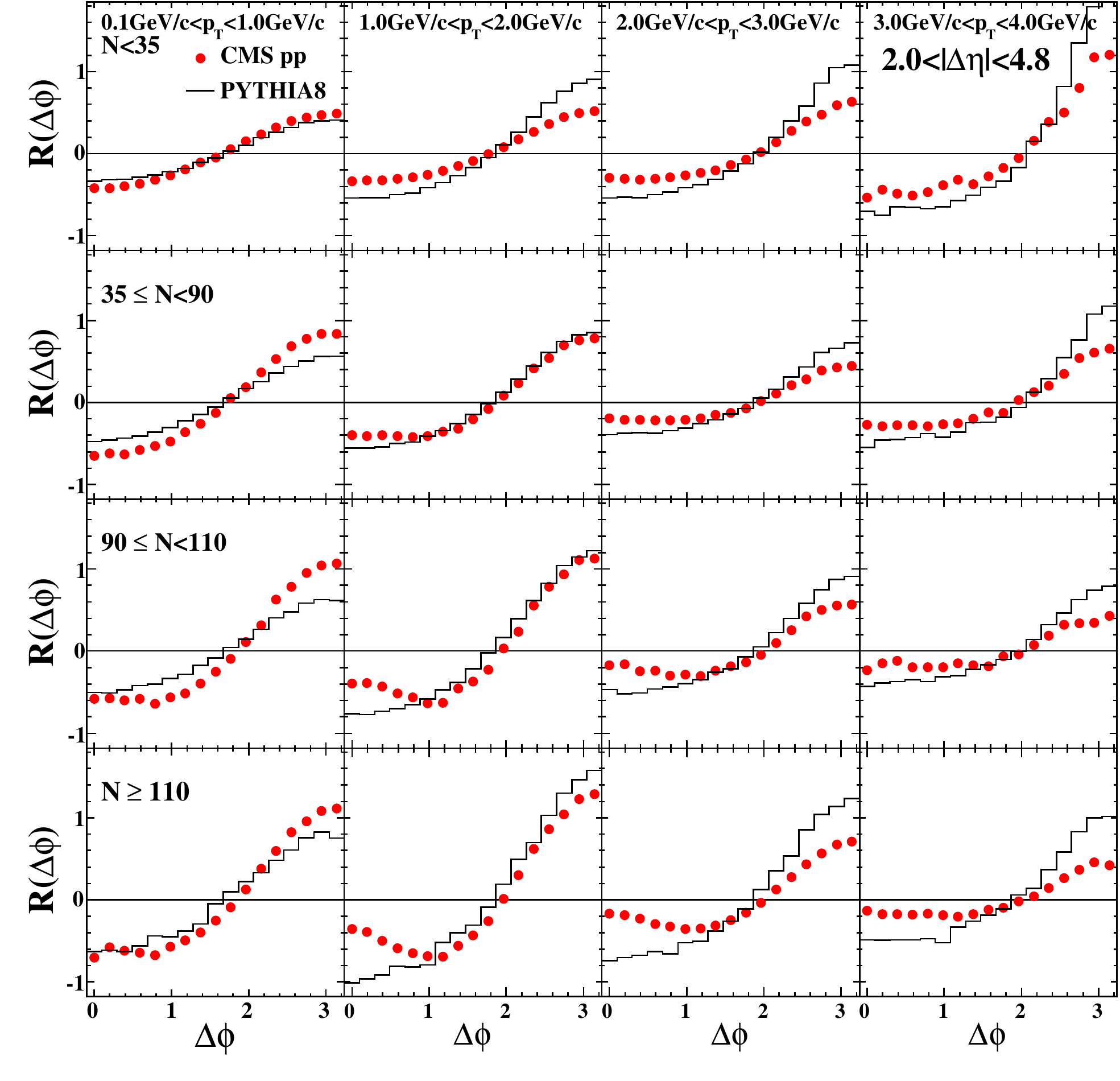}}}
\vspace{-0.3cm}
  \caption{ \label{fig:compare_corr_phi_PPData_Minbias_7TeV_dphi25-48_highmult}
Projections of 2-D correlation functions onto $\Delta\phi$ for $2.0<|\Delta\eta|<4.8$ in different
$p_{T}$ and multiplicity bins for fully corrected  7~TeV $pp$ data and reconstructed PYTHIA8 simulations.
Error bars are smaller than the symbols.
   }
\end{figure}

For $p_{T}$-integrated two-particle correlations in high multiplicity events
($N_\mathrm{trk}^\mathrm{offline} \geq 110$, Fig.~\ref{fig:compare_corr_2D_PPData_Minbias_7TeV_all}c),
most correlation structures are similar to those for minimum bias events.
The cut on high multiplicity enhances the relative contribution of high $p_T$ jets
which fragment into a large number of particles and, therefore, has a qualitatively similar
effect on the shape as the particle $p_T$ cut on minimum bias events
(compare Fig.~\ref{fig:compare_corr_2D_PPData_Minbias_7TeV_all}b
and Fig.~\ref{fig:compare_corr_2D_PPData_Minbias_7TeV_all}c).
However, it is interesting to note that a closer inspection of
the shallow minimum  at $\Delta\phi \approx 0$ and $|\Delta\eta| > 2$
in high multiplicity $p_{T}$-integrated events reveals it to be
slightly less pronounced than that in minimum bias collisions.

Moving to the intermediate $p_{T}$ range in high multiplicity events shown
in Fig.~\ref{fig:compare_corr_2D_PPData_Minbias_7TeV_all}d,
an unexpected effect is observed in the data.
A clear and significant ``ridge''-like structure emerges at $\Delta\phi \approx 0$
extending to $|\Delta\eta|$ of at least 4 units. This is a novel feature
of the data which has never been
seen in two-particle correlation functions in $pp$ or $p\bar{p}$ collisions.
Simulations using MC models do not predict such an effect.
An identical analysis of high multiplicity
events in PYTHIA8~\cite{Sjostrand:2007gs} results in correlation functions which do not exhibit the
extended ridge at $\Delta\phi \approx $0 seen in Fig.~\ref{fig:compare_corr_2D_PPData_Minbias_7TeV_all}d,
while all other structures of the correlation function are qualitatively reproduced. PYTHIA8 was used
to compare to these data since it produces more high multiplicity events than PYTHIA6 in the D6T tune . Several other
PYTHIA tunes, as well as HERWIG++~\cite{Bahr:2008pv}  and Madgraph~\cite{Alwall:2007st} events
were also investigated. No evidence for near-side correlations corresponding to those
seen in data was found.

The novel structure in the high multiplicity $pp$ data is reminiscent
of correlations seen in relativistic heavy ion data.
In the latter case, the observed long-range correlations are generally assumed to arise from various
components of hydrodynamic flow of the produced medium
~\cite{Kolb:2000fha,Back:2004je,Adams:2005dq,Adcox:2004mh,Abelev:2009jv},
from interactions between hard scattering processes and the medium, and from
collective effects in the initial interaction of the nuclei.

However, new correlations can also start to emerge in the  new energy regime
probed here due to more elementary processes.
For example, long range correlations are predicted also to occur in systems
with a large number of fluctuating components, e.g. originating from
additional color string connections. Such effects are
presently not modeled in the MC generators.

Compared to the minimum bias analysis, the online and offline event selection of the rare
high multiplicity events eliminated some sources of systematic uncertainties, but
also introduced several additional ones.  The bias due to the selection efficiency for NSD events, and its associated
correction, were not an issue for the high multiplicity analysis since the efficiency reaches 100\% as discussed in
Section~\ref{sec:inclusive_evtcorr}.
However, it was necessary to correct for the inefficiency in the HLT selection
shown in Fig.~\ref{fig:Multiplicity_ALL}.
Comparison of correlation functions for the high multiplicity bin,
$N_\mathrm{trk}^\mathrm{offline} \geq 110$, taken from the two different trigger paths,
($N_\mathrm{trk}^\mathrm{online}>70$ and $N_\mathrm{trk}^\mathrm{online}>85$, see
Section~\ref{sec:highmult_evtcorr})
showed a systematic variation of 4 to 5\%.

The pile-up rate (fraction of events with more than one good offline vertex found) reached about 40\%
for the LHC conditions pertaining for  most of the high multiplicity data-taking.
Studies on correlations between two offline vertices in each event showed that about 10\% of the events
contained pile-up that could not be distinguished by the vertex finding algorithm.
Therefore, a high multiplicity event could be faked by a pile-up of several minimum bias
collisions with very close vertex positions.
Although such pile-up of
independent $pp$ collisions is not expected to generate additional correlations
in this analysis, a data-driven limit on the effect of pile-up events was established.
This was based on a comparison of  results from runs with negligible
pile-up collected with lower instantaneous luminosity to results obtained with high luminosity
data at nominal bunch intensity. This comparison was limited by the size of the event sample
collected for low luminosity conditions in early LHC running.
For all luminosity selections, the near-side ridge signal
was observed and a conservative systematic error of 15\%, which covers the difference over all
run periods, was assigned.

\begin{figure}[thb]
\centerline{
\hspace{-1.0cm}
\vspace{0.1cm}
  \mbox{\includegraphics[width=\linewidth]{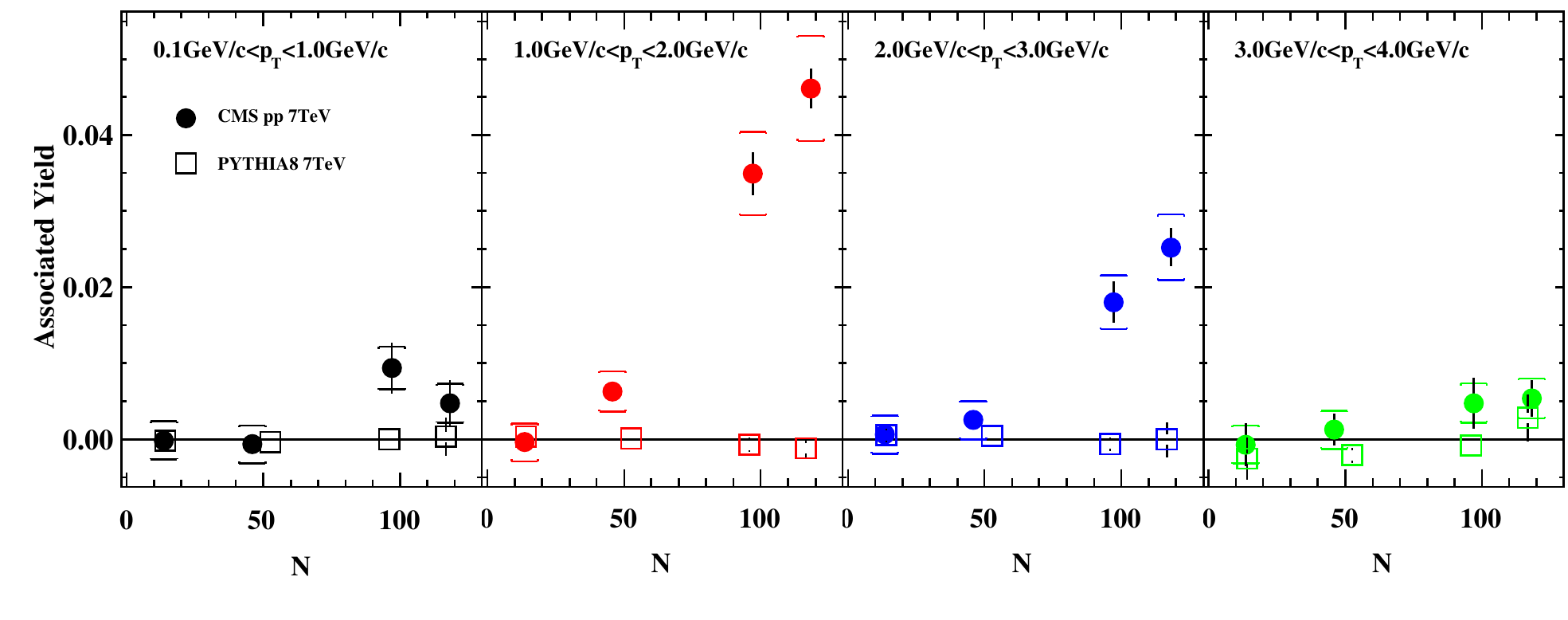}}}
\vspace{-0.3cm}
  \caption{ \label{fig:yieldvsn_PPData_Minbias_7TeV_dphi2-5} Associated yield for the near-side
  of the correlation function integrated over the region of $2.0<|\Delta\eta|<4.8$ as a function
  of event multiplicity in bins of $p_{T}$ for  7~TeV $pp$ collisions.
The error bars correspond to statistical errors, while the brackets around the
data points denote the systematic uncertainties. The open squares show results for PYTHIA8.
   }
\end{figure}

In order to investigate the turn-on behavior of the "ridge"-like structure quantitatively and
in finer detail, correlation functions were obtained in four bins of charged particle
multiplicity and four bins of particle transverse momentum. To study the long-range azimuthal
correlations, the 1-D $\Delta\phi$ were calculated by integrating over the $2.0<|\Delta\eta|<4.8$ region.
Figure~\ref{fig:compare_corr_phi_PPData_Minbias_7TeV_dphi25-48_highmult} shows the results
for a range of $p_{T}$ (from left to right) and multiplicity (from top to bottom) bins. CMS data
are shown as solid circles and the lines show PYTHIA8 results. In this projection, only the
range of $0 < \Delta\phi < \pi$ is shown, as the $\Delta\phi$ correlation function is symmetric around
$\Delta\phi = 0$ by construction. All panels show the away-side jet contribution at $\Delta\phi \approx \pi$.
In addition, for high multiplicity bins in the intermediate $\pt$ region, $1< \pt < 3$~\GeVc,
a second local maximum  near $\Delta\phi \approx 0$ is clearly observed. This new feature of the
long-range azimuthal correlation function is not present in low multiplicity or
minimum bias data, which are dominated by the low multiplicity events.

The comparison of data to PYTHIA8 simulations is characterized by two discrepancies: The strength of
the away-side correlation is over- or underpredicted for almost all bins.
This quantitative discrepancy could be remedied by further
tuning of the relative contributions of di-jet and multi-jet processes compared to particle
production from soft processes in the model without introducing a qualitatively new mechanism.
More importantly, PYTHIA8 qualitatively fails to reproduce the novel local maximum near $\Delta\phi \approx 0$
in any of the $p_T$ or multiplicity bins. It appears that soft particle production from
string fragmentation, the contribution from jet fragmentation, final-state radiation, and
concurrent semihard multi-parton interactions,
to the extent they are parametrized in PYTHIA8, do not provide a mechanism to create the
observed long-range, near-side particle correlations.

\begin{figure}[thb]
\centerline{
\hspace{-1.0cm}
\vspace{0.1cm}
  \mbox{\includegraphics[width=\linewidth]{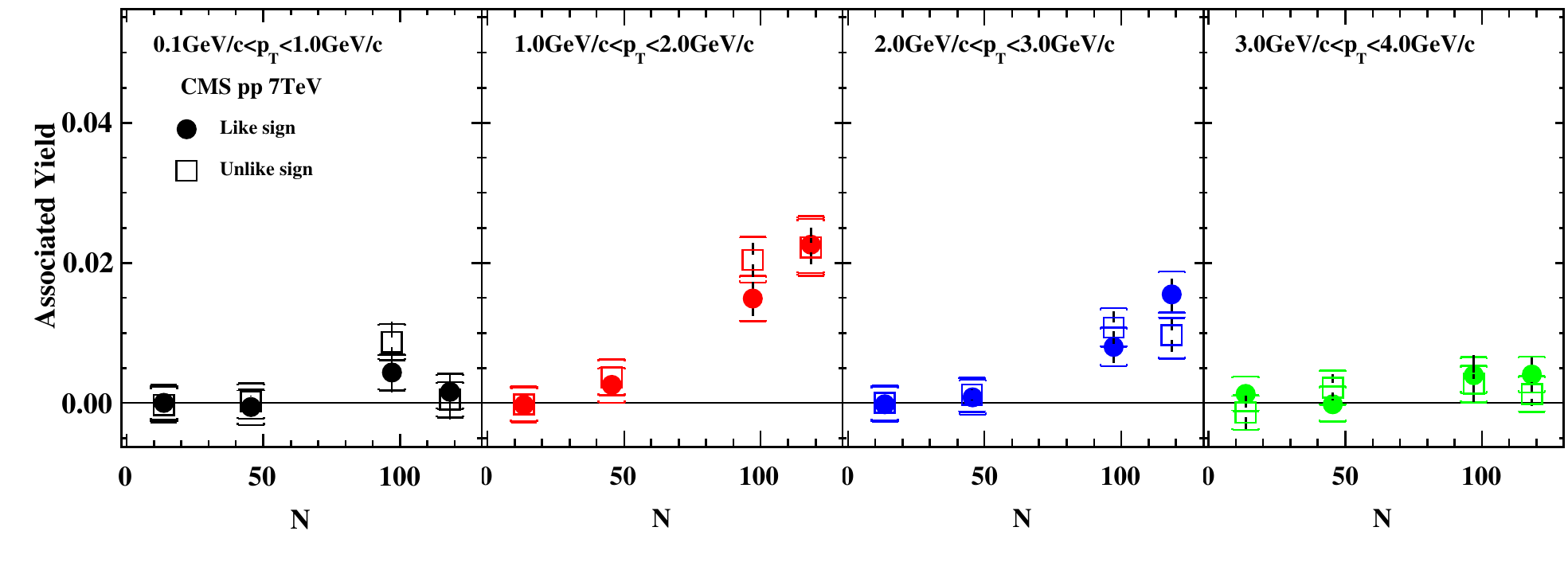}}}
\vspace{-0.3cm}
  \caption{ \label{fig:yieldvsn_PPData_Minbias_7TeV_dphi2-5_sign}
  Like-sign and unlike-sign associated yield for the near-side
  of the correlation function integrated over the region of $2.0<|\Delta\eta|<4.8$ as a function
  of event multiplicity in bins of $p_{T}$.
The error bars correspond to statistical errors, while the brackets around the
data points denote the systematic uncertainties.
   }
\end{figure}

Figure~\ref{fig:compare_corr_phi_PPData_Minbias_7TeV_dphi25-48_highmult} shows that the long-range,
near-side correlation increases in strength with increasing multiplicity and is most prominent
in the region of $1 < \pt < 3$~\GeVc.
The strength of the near-side ridge and its dependence on $p_T$ and multiplicity can
be quantified in more detail by calculating the associated yield, i.e., the number of
other particles correlated with a specific particle. In the presence of multiple sources of correlations,
the yield for the correlation of interest is commonly estimated using an implementation of the
zero-yield-at-minimum (ZYAM) method~\cite{PhysRevC.72.011902}.
The procedure uses R($\Delta\phi$)
integrated over $2.0<|\Delta\eta|<4.8$ as shown in Fig.~\ref{fig:compare_corr_phi_PPData_Minbias_7TeV_dphi25-48_highmult}.
In the first step,
a second-order polynomial is fit to R($\Delta\phi$) in the region $0.1<|\Delta\phi|<2.0$. The location
of the minimum of the polynomial in this region is  denoted $\Delta\phi_{\rm ZYAM}$.
The contribution from the background
source of correlations, in this case the away-side jet correlations, is assumed to be zero for
$ |\Delta\phi| \leq \Delta\phi_{\rm ZYAM}$.
Using the position of the minimum, the associated yield is then found by integrating
R($\Delta\phi$) over the region $0<|\Delta\phi|<\Delta\phi_{\rm ZYAM}$ relative to the
minimum in R($\Delta\phi$) and multiplying by $\int_{2.0}^{4.8}$B($\Delta\eta$)d$|\Delta\eta|$ to
account for the fact that only a limited $\Delta\eta$ range is used.
The uncertainty on the minimum level of $R(\Delta\phi)$ obtained by the ZYAM procedure
as well as varying the fit range in $\Delta\phi$ gives an
uncertainty of 0.0025 on the associated yield, uniformly over all multiplicity
and $p_T$ bins.

Figure~\ref{fig:yieldvsn_PPData_Minbias_7TeV_dphi2-5} shows the associated yield as a function
of event multiplicity integrated over $2.0<|\Delta\eta|<4.8$ in increasing bins of $p_{T}$.
The ridge yield is consistent with zero for low multiplicity events. The emergence of the ridge is observed toward
the very high multiplicity region, primarily for the intermediate $p_{T}$ range of $1-3\GeVc$.
The error bars correspond to statistical errors, while the brackets around the data points denote
the systematic uncertainties.
Results from the PYTHIA8 MC, shown in open squares in
Fig.~\ref{fig:yieldvsn_PPData_Minbias_7TeV_dphi2-5}, are consistent with zero for all multiplicity
and $p_{T}$ regions, indicating that the ridge observed in the data is totally absent in events
produced by this generator.

To investigate the novel ridge-like structure further, two-particle correlations were
calculated separately for like-sign and unlike-sign charged pairs.
Possible problems related to the track reconstruction algorithm,
like multiple reconstruction of the same particle
or local occupancy changes, would be expected to affect like-sign pairs differently than
unlike-sign pairs.
The same choice of pairs of like- or unlike-sign was made for both the signal and background
in Eq.~(\ref{2pcorr_incl}).
Figure~\ref{fig:yieldvsn_PPData_Minbias_7TeV_dphi2-5_sign}
shows the associated yield for like-sign (solid circles) and unlike-sign (open squares)
two-particle correlations respectively as a function of event multiplicity integrated over
$2.0<|\Delta\eta|<4.8$ in bins of $p_{T}$.
Consistent multiplicity and $\pt$ dependencies
of the near-side associated yield are observed for charge dependent and charge independent correlations.
The results for like-sign and unlike-sign pairs agree with each other within uncertainties.
Since the number of like- and unlike-sign pairs each represent roughly half of the total, the
yield of associated pairs counting only one sign option is expected to be roughly a factor of
two smaller than the unrestricted yield.

As a further cross-check, correlation functions were generated for tracks paired with ECAL
photons (primarily due to $\pi^0$s) as well as pairs of two ECAL photons. These
distributions showed similar behavior to those shown in
Figs.~\ref{fig:compare_corr_2D_PPData_Minbias_7TeV_all} and
\ref{fig:compare_corr_phi_PPData_Minbias_7TeV_dphi25-48_highmult}, i.e., the high
$|\Delta\eta|$ region contained a dip at $|\Delta\phi| \approx 0$ in minimum bias events and a
ridge in that region for high multiplicity events. Data at 0.9 and 2.36~TeV were also
analyzed for long-range correlations, but the statistics were not sufficient to draw a conclusion.

\section{Conclusion}
\label{sec:conclusion}

The CMS detector at the LHC has been used to measure angular correlations
between two charged particles up to $|\Delta \eta|\approx 5$ and over the full range of
$\Delta \phi$ in $pp$ collisions at $\sqrt{s}$ = 0.9, 2.36, and 7~TeV.
The extracted 2-D correlation functions show a variety of features. In minimum
bias collisions they are dominated by a maximum at $\Delta \eta = 0$ extending
over the full range in $\Delta \phi$, with a width which tends to increase with increasing $\Delta \phi$.
A simple cluster model parametrization was fit to these short-range
correlations in order to
quantify their strength (the effective cluster size) and their
extent in relative pseudorapidity (the cluster decay width).
The cluster size is observed to increase slowly with beam energy,
while the cluster width is essentially constant.
The PYTHIA event generator with D6T tune correctly describes the cluster
widths and the energy dependence of the cluster size
but systematically underestimates the cluster size.

Long-range azimuthal correlations for $2.0 < |\Delta\eta| < 4.8$ have been
studied for 7~TeV data, leading to the first observation of a long-range
ridge-like structure at the near-side ($\Delta\phi \approx 0$) in $pp$ collisions. This striking feature
is clearly seen for large rapidity differences $|\Delta\eta| > 2$ in
events with an observed charged particle multiplicity of $N \approx 90$ or higher.
The enhancement in the near-side correlation function is most evident
in the intermediate transverse momentum range, $1 < p_{T} < 3$~\GeVc.
In the $2.0 < |\Delta\eta| < 4.8$ range, a steep increase
of the near-side associated yield with multiplicity has been found in the data, whereas
simulations show an associated yield consistent with zero, independent of
multiplicity and transverse momentum.
The novel structure resembles similar features observed in heavy
ion experiments~\cite{Alver:2008gk,Alver:2009id,Abelev:2009jv}.
However, the physical origin of our observation is not yet understood.
Additional characteristics of the high multiplicity $pp$ events
displaying this novel feature deserve further detailed study.

\bibliography{auto_generated}   

\cleardoublepage\appendix\section{The CMS Collaboration \label{app:collab}}\begin{sloppypar}\hyphenpenalty=5000\widowpenalty=500\clubpenalty=5000\input{QCD-10-002-authorlist.tex}\end{sloppypar}
\end{document}

%% file: ptdr-definitions.tex
%
%
%

\providecommand {\etal}{\mbox{et al.}\xspace} 
\providecommand {\ie}{\mbox{i.e.}\xspace}     
\providecommand {\eg}{\mbox{e.g.}\xspace}     
\providecommand {\etc}{\mbox{etc.}\xspace}     
\providecommand {\vs}{\mbox{\sl vs.}\xspace}      
\providecommand {\mdash}{\ensuremath{\mathrm{-}}} 

\providecommand {\Lone}{Level-1\xspace} 
\providecommand {\Ltwo}{Level-2\xspace}
\providecommand {\Lthree}{Level-3\xspace}

\providecommand{\ACERMC} {\textsc{AcerMC}\xspace}
\providecommand{\ALPGEN} {{\textsc{alpgen}}\xspace}
\providecommand{\CHARYBDIS} {{\textsc{charybdis}}\xspace}
\providecommand{\CMKIN} {\textsc{cmkin}\xspace}
\providecommand{\CMSIM} {{\textsc{cmsim}}\xspace}
\providecommand{\CMSSW} {{\textsc{cmssw}}\xspace}
\providecommand{\COBRA} {{\textsc{cobra}}\xspace}
\providecommand{\COCOA} {{\textsc{cocoa}}\xspace}
\providecommand{\COMPHEP} {\textsc{CompHEP}\xspace}
\providecommand{\EVTGEN} {{\textsc{evtgen}}\xspace}
\providecommand{\FAMOS} {{\textsc{famos}}\xspace}
\providecommand{\GARCON} {\textsc{garcon}\xspace}
\providecommand{\GARFIELD} {{\textsc{garfield}}\xspace}
\providecommand{\GEANE} {{\textsc{geane}}\xspace}
\providecommand{\GEANTfour} {{\textsc{geant4}}\xspace}
\providecommand{\GEANTthree} {{\textsc{geant3}}\xspace}
\providecommand{\GEANT} {{\textsc{geant}}\xspace}
\providecommand{\HDECAY} {\textsc{hdecay}\xspace}
\providecommand{\HERWIG} {{\textsc{herwig}}\xspace}
\providecommand{\HIGLU} {{\textsc{higlu}}\xspace}
\providecommand{\HIJING} {{\textsc{hijing}}\xspace}
\providecommand{\IGUANA} {\textsc{iguana}\xspace}
\providecommand{\ISAJET} {{\textsc{isajet}}\xspace}
\providecommand{\ISAPYTHIA} {{\textsc{isapythia}}\xspace}
\providecommand{\ISASUGRA} {{\textsc{isasugra}}\xspace}
\providecommand{\ISASUSY} {{\textsc{isasusy}}\xspace}
\providecommand{\ISAWIG} {{\textsc{isawig}}\xspace}
\providecommand{\MADGRAPH} {\textsc{MadGraph}\xspace}
\providecommand{\MCATNLO} {\textsc{mc@nlo}\xspace}
\providecommand{\MCFM} {\textsc{mcfm}\xspace}
\providecommand{\MILLEPEDE} {{\textsc{millepede}}\xspace}
\providecommand{\ORCA} {{\textsc{orca}}\xspace}
\providecommand{\OSCAR} {{\textsc{oscar}}\xspace}
\providecommand{\PHOTOS} {\textsc{photos}\xspace}
\providecommand{\PROSPINO} {\textsc{prospino}\xspace}
\providecommand{\PYTHIA} {{\textsc{pythia}}\xspace}
\providecommand{\SHERPA} {{\textsc{sherpa}}\xspace}
\providecommand{\TAUOLA} {\textsc{tauola}\xspace}
\providecommand{\TOPREX} {\textsc{TopReX}\xspace}
\providecommand{\XDAQ} {{\textsc{xdaq}}\xspace}

\providecommand {\DZERO}{D\O\xspace}     


\providecommand{\de}{\ensuremath{^\circ}}
\providecommand{\ten}[1]{\ensuremath{\times \text{10}^\text{#1}}}
\providecommand{\unit}[1]{\ensuremath{\text{\,#1}}\xspace}
\providecommand{\mum}{\ensuremath{\,\mu\text{m}}\xspace}
\providecommand{\micron}{\ensuremath{\,\mu\text{m}}\xspace}
\providecommand{\cm}{\ensuremath{\,\text{cm}}\xspace}
\providecommand{\mm}{\ensuremath{\,\text{mm}}\xspace}
\providecommand{\mus}{\ensuremath{\,\mu\text{s}}\xspace}
\providecommand{\keV}{\ensuremath{\,\text{ke\hspace{-.08em}V}}\xspace}
\providecommand{\MeV}{\ensuremath{\,\text{Me\hspace{-.08em}V}}\xspace}
\providecommand{\GeV}{\ensuremath{\,\text{Ge\hspace{-.08em}V}}\xspace}
\providecommand{\TeV}{\ensuremath{\,\text{Te\hspace{-.08em}V}}\xspace}
\providecommand{\PeV}{\ensuremath{\,\text{Pe\hspace{-.08em}V}}\xspace}
\providecommand{\keVc}{\ensuremath{{\,\text{ke\hspace{-.08em}V\hspace{-0.16em}/\hspace{-0.08em}}c}}\xspace}
\providecommand{\MeVc}{\ensuremath{{\,\text{Me\hspace{-.08em}V\hspace{-0.16em}/\hspace{-0.08em}}c}}\xspace}
\providecommand{\GeVc}{\ensuremath{{\,\text{Ge\hspace{-.08em}V\hspace{-0.16em}/\hspace{-0.08em}}c}}\xspace}
\providecommand{\TeVc}{\ensuremath{{\,\text{Te\hspace{-.08em}V\hspace{-0.16em}/\hspace{-0.08em}}c}}\xspace}
\providecommand{\keVcc}{\ensuremath{{\,\text{ke\hspace{-.08em}V\hspace{-0.16em}/\hspace{-0.08em}}c^\text{2}}}\xspace}
\providecommand{\MeVcc}{\ensuremath{{\,\text{Me\hspace{-.08em}V\hspace{-0.16em}/\hspace{-0.08em}}c^\text{2}}}\xspace}
\providecommand{\GeVcc}{\ensuremath{{\,\text{Ge\hspace{-.08em}V\hspace{-0.16em}/\hspace{-0.08em}}c^\text{2}}}\xspace}
\providecommand{\TeVcc}{\ensuremath{{\,\text{Te\hspace{-.08em}V\hspace{-0.16em}/\hspace{-0.08em}}c^\text{2}}}\xspace}

\providecommand{\pbinv} {\mbox{\ensuremath{\,\text{pb}^\text{$-$1}}}\xspace}
\providecommand{\fbinv} {\mbox{\ensuremath{\,\text{fb}^\text{$-$1}}}\xspace}
\providecommand{\nbinv} {\mbox{\ensuremath{\,\text{nb}^\text{$-$1}}}\xspace}
\providecommand{\percms}{\ensuremath{\,\text{cm}^\text{$-$2}\,\text{s}^\text{$-$1}}\xspace}
\providecommand{\lumi}{\ensuremath{\mathcal{L}}\xspace}
\providecommand{\Lumi}{\ensuremath{\mathcal{L}}\xspace}
%
\providecommand{\LvLow}  {\ensuremath{\mathcal{L}=\text{10}^\text{32}\,\text{cm}^\text{$-$2}\,\text{s}^\text{$-$1}}\xspace}
\providecommand{\LLow}   {\ensuremath{\mathcal{L}=\text{10}^\text{33}\,\text{cm}^\text{$-$2}\,\text{s}^\text{$-$1}}\xspace}
\providecommand{\lowlumi}{\ensuremath{\mathcal{L}=\text{2}\times \text{10}^\text{33}\,\text{cm}^\text{$-$2}\,\text{s}^\text{$-$1}}\xspace}
\providecommand{\LMed}   {\ensuremath{\mathcal{L}=\text{2}\times \text{10}^\text{33}\,\text{cm}^\text{$-$2}\,\text{s}^\text{$-$1}}\xspace}
\providecommand{\LHigh}  {\ensuremath{\mathcal{L}=\text{10}^\text{34}\,\text{cm}^\text{$-$2}\,\text{s}^\text{$-$1}}\xspace}
\providecommand{\hilumi} {\ensuremath{\mathcal{L}=\text{10}^\text{34}\,\text{cm}^\text{$-$2}\,\text{s}^\text{$-$1}}\xspace}


\providecommand{\zp}{\ensuremath{\mathrm{Z}^\prime}\xspace}


\providecommand{\kt}{\ensuremath{k_{\mathrm{T}}}\xspace}
\providecommand{\BC}{\ensuremath{\mathrm{B_{c}}}\xspace}
\providecommand{\bbarc}{\ensuremath{\mathrm{\overline{b}c}}\xspace}
\providecommand{\bbbar}{\ensuremath{\mathrm{b\overline{b}}}\xspace}
\providecommand{\ccbar}{\ensuremath{\mathrm{c\overline{c}}}\xspace}
\providecommand{\JPsi}{\ensuremath{\mathrm{J}\hspace{-.08em}/\hspace{-.14em}\psi}\xspace}
\providecommand{\bspsiphi}{\ensuremath{\mathrm{B_s} \to \JPsi\, \phi}\xspace}
\providecommand{\AFB}{\ensuremath{A_\text{FB}}\xspace}
\providecommand{\EE}{\ensuremath{\mathrm{e^+e^-}}\xspace}
\providecommand{\MM}{\ensuremath{\mu^+\mu^-}\xspace}
\providecommand{\TT}{\ensuremath{\tau^+\tau^-}\xspace}
\providecommand{\wangle}{\ensuremath{\sin^{2}\theta_{\text{eff}}^\text{lept}(M^2_\mathrm{Z})}\xspace}
\providecommand{\ttbar}{\ensuremath{\mathrm{t\overline{t}}}\xspace}
\providecommand{\stat}{\ensuremath{\,\text{(stat.)}}\xspace}
\providecommand{\syst}{\ensuremath{\,\text{(syst.)}}\xspace}

\providecommand{\HGG}{\ensuremath{\mathrm{H}\to\gamma\gamma}}
\providecommand{\gev}{\GeV}
\providecommand{\GAMJET}{\ensuremath{\gamma + \text{jet}}}
\providecommand{\PPTOJETS}{\ensuremath{\mathrm{pp}\to\text{jets}}}
\providecommand{\PPTOGG}{\ensuremath{\mathrm{pp}\to\gamma\gamma}}
\providecommand{\PPTOGAMJET}{\ensuremath{\mathrm{pp}\to\gamma + \mathrm{jet}}}
\providecommand{\MH}{\ensuremath{M_{\mathrm{H}}}}
\providecommand{\RNINE}{\ensuremath{R_\mathrm{9}}}
\providecommand{\DR}{\ensuremath{\Delta R}}


\providecommand{\PT}{\ensuremath{p_{\mathrm{T}}}\xspace}
\providecommand{\pt}{\ensuremath{p_{\mathrm{T}}}\xspace}
\providecommand{\ET}{\ensuremath{E_{\mathrm{T}}}\xspace}
\providecommand{\HT}{\ensuremath{H_{\mathrm{T}}}\xspace}
\providecommand{\et}{\ensuremath{E_{\mathrm{T}}}\xspace}
\providecommand{\Em}{\ensuremath{E\hspace{-0.6em}/}\xspace}
\providecommand{\Pm}{\ensuremath{p\hspace{-0.5em}/}\xspace}
\providecommand{\PTm}{\ensuremath{{p}_\mathrm{T}\hspace{-1.02em}/}\xspace}
\providecommand{\PTslash}{\ensuremath{{p}_\mathrm{T}\hspace{-1.02em}/}\xspace}
\providecommand{\ETm}{\ensuremath{E_{\mathrm{T}}^{\text{miss}}}\xspace}
\providecommand{\ETslash}{\ensuremath{E_{\mathrm{T}}\hspace{-1.1em}/}\xspace}
\providecommand{\MET}{\ensuremath{E_{\mathrm{T}}^{\text{miss}}}\xspace}
\providecommand{\ETmiss}{\ensuremath{E_{\mathrm{T}}^{\text{miss}}}\xspace}
\providecommand{\VEtmiss}{\ensuremath{{\vec E}_{\mathrm{T}}^{\text{miss}}}\xspace}

\providecommand{\dd}[2]{\ensuremath{\frac{\mathrm{d} #1}{\mathrm{d} #2}}}

%

\providecommand{\ga}{\ensuremath{\gtrsim}}
\providecommand{\la}{\ensuremath{\lesssim}}
\providecommand{\swsq}{\ensuremath{\sin^2\theta_\mathrm{W}}\xspace}
\providecommand{\cwsq}{\ensuremath{\cos^2\theta_\mathrm{W}}\xspace}
\providecommand{\tanb}{\ensuremath{\tan\beta}\xspace}
\providecommand{\tanbsq}{\ensuremath{\tan^{2}\beta}\xspace}
\providecommand{\sidb}{\ensuremath{\sin 2\beta}\xspace}
\providecommand{\alpS}{\ensuremath{\alpha_S}\xspace}
\providecommand{\alpt}{\ensuremath{\tilde{\alpha}}\xspace}

\providecommand{\QL}{\ensuremath{\mathrm{Q}_\mathrm{L}}\xspace}
\providecommand{\sQ}{\ensuremath{\tilde{\mathrm{Q}}}\xspace}
\providecommand{\sQL}{\ensuremath{\tilde{\mathrm{Q}}_\mathrm{L}}\xspace}
\providecommand{\ULC}{\ensuremath{\mathrm{U}_\mathrm{L}^\mathrm{C}}\xspace}
\providecommand{\sUC}{\ensuremath{\tilde{\mathrm{U}}^\mathrm{C}}\xspace}
\providecommand{\sULC}{\ensuremath{\tilde{\mathrm{U}}_\mathrm{L}^\mathrm{C}}\xspace}
\providecommand{\DLC}{\ensuremath{\mathrm{D}_\mathrm{L}^\mathrm{C}}\xspace}
\providecommand{\sDC}{\ensuremath{\tilde{\mathrm{D}}^\mathrm{C}}\xspace}
\providecommand{\sDLC}{\ensuremath{\tilde{\mathrm{D}}_\mathrm{L}^\mathrm{C}}\xspace}
\providecommand{\LL}{\ensuremath{\mathrm{L}_\mathrm{L}}\xspace}
\providecommand{\sL}{\ensuremath{\tilde{\mathrm{L}}}\xspace}
\providecommand{\sLL}{\ensuremath{\tilde{\mathrm{L}}_\mathrm{L}}\xspace}
\providecommand{\ELC}{\ensuremath{\mathrm{E}_\mathrm{L}^\mathrm{C}}\xspace}
\providecommand{\sEC}{\ensuremath{\tilde{\mathrm{E}}^\mathrm{C}}\xspace}
\providecommand{\sELC}{\ensuremath{\tilde{\mathrm{E}}_\mathrm{L}^\mathrm{C}}\xspace}
\providecommand{\sEL}{\ensuremath{\tilde{\mathrm{E}}_\mathrm{L}}\xspace}
\providecommand{\sER}{\ensuremath{\tilde{\mathrm{E}}_\mathrm{R}}\xspace}
\providecommand{\sFer}{\ensuremath{\tilde{\mathrm{f}}}\xspace}
\providecommand{\sQua}{\ensuremath{\tilde{\mathrm{q}}}\xspace}
\providecommand{\sUp}{\ensuremath{\tilde{\mathrm{u}}}\xspace}
\providecommand{\suL}{\ensuremath{\tilde{\mathrm{u}}_\mathrm{L}}\xspace}
\providecommand{\suR}{\ensuremath{\tilde{\mathrm{u}}_\mathrm{R}}\xspace}
\providecommand{\sDw}{\ensuremath{\tilde{\mathrm{d}}}\xspace}
\providecommand{\sdL}{\ensuremath{\tilde{\mathrm{d}}_\mathrm{L}}\xspace}
\providecommand{\sdR}{\ensuremath{\tilde{\mathrm{d}}_\mathrm{R}}\xspace}
\providecommand{\sTop}{\ensuremath{\tilde{\mathrm{t}}}\xspace}
\providecommand{\stL}{\ensuremath{\tilde{\mathrm{t}}_\mathrm{L}}\xspace}
\providecommand{\stR}{\ensuremath{\tilde{\mathrm{t}}_\mathrm{R}}\xspace}
\providecommand{\stone}{\ensuremath{\tilde{\mathrm{t}}_1}\xspace}
\providecommand{\sttwo}{\ensuremath{\tilde{\mathrm{t}}_2}\xspace}
\providecommand{\sBot}{\ensuremath{\tilde{\mathrm{b}}}\xspace}
\providecommand{\sbL}{\ensuremath{\tilde{\mathrm{b}}_\mathrm{L}}\xspace}
\providecommand{\sbR}{\ensuremath{\tilde{\mathrm{b}}_\mathrm{R}}\xspace}
\providecommand{\sbone}{\ensuremath{\tilde{\mathrm{b}}_1}\xspace}
\providecommand{\sbtwo}{\ensuremath{\tilde{\mathrm{b}}_2}\xspace}
\providecommand{\sLep}{\ensuremath{\tilde{\mathrm{l}}}\xspace}
\providecommand{\sLepC}{\ensuremath{\tilde{\mathrm{l}}^\mathrm{C}}\xspace}
\providecommand{\sEl}{\ensuremath{\tilde{\mathrm{e}}}\xspace}
\providecommand{\sElC}{\ensuremath{\tilde{\mathrm{e}}^\mathrm{C}}\xspace}
\providecommand{\seL}{\ensuremath{\tilde{\mathrm{e}}_\mathrm{L}}\xspace}
\providecommand{\seR}{\ensuremath{\tilde{\mathrm{e}}_\mathrm{R}}\xspace}
\providecommand{\snL}{\ensuremath{\tilde{\nu}_L}\xspace}
\providecommand{\sMu}{\ensuremath{\tilde{\mu}}\xspace}
\providecommand{\sNu}{\ensuremath{\tilde{\nu}}\xspace}
\providecommand{\sTau}{\ensuremath{\tilde{\tau}}\xspace}
\providecommand{\Glu}{\ensuremath{\mathrm{g}}\xspace}
\providecommand{\sGlu}{\ensuremath{\tilde{\mathrm{g}}}\xspace}
\providecommand{\Wpm}{\ensuremath{\mathrm{W}^{\pm}}\xspace}
\providecommand{\sWpm}{\ensuremath{\tilde{\mathrm{W}}^{\pm}}\xspace}
\providecommand{\Wz}{\ensuremath{\mathrm{W}^{0}}\xspace}
\providecommand{\sWz}{\ensuremath{\tilde{\mathrm{W}}^{0}}\xspace}
\providecommand{\sWino}{\ensuremath{\tilde{\mathrm{W}}}\xspace}
\providecommand{\Bz}{\ensuremath{\mathrm{B}^{0}}\xspace}
\providecommand{\sBz}{\ensuremath{\tilde{\mathrm{B}}^{0}}\xspace}
\providecommand{\sBino}{\ensuremath{\tilde{\mathrm{B}}}\xspace}
\providecommand{\Zz}{\ensuremath{\mathrm{Z}^{0}}\xspace}
\providecommand{\sZino}{\ensuremath{\tilde{\mathrm{Z}}^{0}}\xspace}
\providecommand{\sGam}{\ensuremath{\tilde{\gamma}}\xspace}
\providecommand{\chiz}{\ensuremath{\tilde{\chi}^{0}}\xspace}
\providecommand{\chip}{\ensuremath{\tilde{\chi}^{+}}\xspace}
\providecommand{\chim}{\ensuremath{\tilde{\chi}^{-}}\xspace}
\providecommand{\chipm}{\ensuremath{\tilde{\chi}^{\pm}}\xspace}
\providecommand{\Hone}{\ensuremath{\mathrm{H}_\mathrm{d}}\xspace}
\providecommand{\sHone}{\ensuremath{\tilde{\mathrm{H}}_\mathrm{d}}\xspace}
\providecommand{\Htwo}{\ensuremath{\mathrm{H}_\mathrm{u}}\xspace}
\providecommand{\sHtwo}{\ensuremath{\tilde{\mathrm{H}}_\mathrm{u}}\xspace}
\providecommand{\sHig}{\ensuremath{\tilde{\mathrm{H}}}\xspace}
\providecommand{\sHa}{\ensuremath{\tilde{\mathrm{H}}_\mathrm{a}}\xspace}
\providecommand{\sHb}{\ensuremath{\tilde{\mathrm{H}}_\mathrm{b}}\xspace}
\providecommand{\sHpm}{\ensuremath{\tilde{\mathrm{H}}^{\pm}}\xspace}
\providecommand{\hz}{\ensuremath{\mathrm{h}^{0}}\xspace}
\providecommand{\Hz}{\ensuremath{\mathrm{H}^{0}}\xspace}
\providecommand{\Az}{\ensuremath{\mathrm{A}^{0}}\xspace}
\providecommand{\Hpm}{\ensuremath{\mathrm{H}^{\pm}}\xspace}
\providecommand{\sGra}{\ensuremath{\tilde{\mathrm{G}}}\xspace}
\providecommand{\mtil}{\ensuremath{\tilde{m}}\xspace}
\providecommand{\rpv}{\ensuremath{\rlap{\kern.2em/}R}\xspace}
\providecommand{\LLE}{\ensuremath{LL\bar{E}}\xspace}
\providecommand{\LQD}{\ensuremath{LQ\bar{D}}\xspace}
\providecommand{\UDD}{\ensuremath{\overline{UDD}}\xspace}
\providecommand{\Lam}{\ensuremath{\lambda}\xspace}
\providecommand{\Lamp}{\ensuremath{\lambda'}\xspace}
\providecommand{\Lampp}{\ensuremath{\lambda''}\xspace}
\providecommand{\spinbd}[2]{\ensuremath{\bar{#1}_{\dot{#2}}}\xspace}

\providecommand{\MD}{\ensuremath{{M_\mathrm{D}}}\xspace}
\providecommand{\Mpl}{\ensuremath{{M_\mathrm{Pl}}}\xspace}
\providecommand{\Rinv} {\ensuremath{{R}^{-1}}\xspace} 

%% file: QCD-10-002-authorlist.tex
\textbf{Yerevan Physics Institute,  Yerevan,  Armenia}\\*[0pt]
V.~Khachatryan, A.M.~Sirunyan, A.~Tumasyan
\vskip\cmsinstskip
\textbf{Institut f\"{u}r Hochenergiephysik der OeAW,  Wien,  Austria}\\*[0pt]
W.~Adam, T.~Bergauer, M.~Dragicevic, J.~Er\"{o}, C.~Fabjan, M.~Friedl, R.~Fr\"{u}hwirth, V.M.~Ghete, J.~Hammer\cmsAuthorMark{1}, S.~H\"{a}nsel, C.~Hartl, M.~Hoch, N.~H\"{o}rmann, J.~Hrubec, M.~Jeitler, G.~Kasieczka, W.~Kiesenhofer, M.~Krammer, D.~Liko, I.~Mikulec, M.~Pernicka, H.~Rohringer, R.~Sch\"{o}fbeck, J.~Strauss, A.~Taurok, F.~Teischinger, W.~Waltenberger, G.~Walzel, E.~Widl, C.-E.~Wulz
% ...
\vskip\cmsinstskip
% ...
1:~~Also at CERN, European Organization for Nuclear Research, Geneva, Switzerland\\
% ...
